\documentclass[a4paper]{article}
\usepackage[margin=1in]{geometry}
\usepackage{graphicx,amsmath,amssymb,mathrsfs,balance,flushend,float,tabularx,array,makecell,dsfont,cancel,enumitem,appendix,url,cite}
\usepackage[english]{babel}

\title{Semiclassical Gravity Beyond General Relativity: Insights from Torsion}

\author{R. Morales-Cabrera and Y. Bonder$^*$\\
$^*$bonder@nucleares.unam.mx\\
Instituto de Ciencias Nucleares\\
Universidad Nacional Aut\'onoma de M\'exico\\
Apartado Postal 70-543, Ciudad de M\'exico, 04510, M\'exico}

\begin{document}

\maketitle

\begin{abstract}
We develop a semiclassical theory of modified gravity with nontrivial spacetime torsion. In particular, we show that the semiclassical treatment can be axiomatized in the case of Einstein--Cartan theory with a nonminimally coupled, free Klein--Gordon field, in four dimensions. Using Hadamard renormalization, we obtain well-defined expectation values for both, the energy--momentum and spin--density operators. These objects exhibit scale and renormalization ambiguities; we identify the latter by constructing a renormalization Lagrangian in terms of differential forms, which are particularly well suited for this purpose. Furthermore, we analyze the conformal anomaly, which persists in the presence of torsion.
\end{abstract}

\maketitle

\section{Introduction}

General Relativity (GR) is the currently accepted theory of gravity. In this framework, gravity is interpreted as the curvature of spacetime, and the spacetime metric, which is the only gravitational field, is determined by Einstein’s equations. In turn, these equations are sourced by the matter energy--momentum tensor.

GR has been experimentally confirmed to a high degree of precision~\cite{Will_2014}. However, several open questions remain, such as the nature of dark matter and dark energy~\cite{Debono_2016}, and the possible resolution of singularities~\cite{Penrose}. Attempts to address these issues often invoke modified gravity theories~\cite{Shankaranarayanan_2022}. Prominent examples are \( f(R) \) models~\cite{fofR}, scalar--tensor theories~\cite{Damour_1992}, and Chern--Simons gravity~\cite{Zanelli:2012px}, as well as frameworks where the spacetime geometry is not fully determined by the metric. The latter includes Einstein--Cartan theory~\cite{Hehl_1976,Hehl}, which incorporates a nontrivial spacetime torsion.

The inclusion of torsion leads to theories with interesting properties. For instance, torsion can modify the singularity theorems~\cite{Luz_2020} and alter the behavior of the early Universe~\cite{NURGALIEV1983378}. Moreover, it allows for an energy--momentum tensor that, in general, is not divergence-free~\cite{Hehl_1976}.

Importantly, the empirical success of GR does not rule out the presence of torsion. In fact, Einstein--Cartan theory reproduces all experimentally tested predictions of GR~\cite{Bonder_2016}. What is more, within the framework of the Standard Model, possible signatures of torsion can be sought as new interactions involving polarized Dirac spinors~\cite{Bruno}.

A fundamental assumption in any theory of gravity, including GR and Einstein--Cartan theory, is that all matter fields are classical. However, matter is most accurately described by quantum mechanics. Indeed, quantum field theory provides some of the most precise experimental validations of any physical theory to date~\cite{Fan_130}. Nevertheless, gravitational interactions have not been consistently incorporated into the quantum framework. This is mostly due to conceptual issues arising from Einstein’s equations: the ``left-hand side'' involves the metric tensor, a classical field, while the ``right-hand side'' depends on matter, described by quantum fields. This inconsistency, together with other considerations~\cite{Carlip}, points to the need for a more fundamental theory, commonly referred to as quantum gravity. Despite extensive efforts, including candidates such as string theory~\cite{Zwiebach:789942} and loop quantum gravity~\cite{Ashtekar_2021}, no fully satisfactory formulation of quantum gravity has been achieved.

A relevant question, therefore, is how to incorporate certain quantum aspects of the matter description into gravity. One approach is to consider quantum field theory in curved spacetime (QFTCS)~\cite{Parker1969a,Parker1971,BD1982,fulling,Wald1994,parkerQFT}. In this framework, the geometry is fixed and quantum matter fields propagate on it. Although QFTCS has provided profound theoretical insights, such as the Unruh effect~\cite{unruh} and Hawking radiation~\cite{Hawking1975}, additional steps are required to account for the influence of quantum matter on spacetime geometry. One possibility is semiclassical gravity~\cite{Moller1962,Rosenfeld1963,Ford}, where gravity is sourced by the expectation values of quantum matter fields. Notably, even though QFTCS does not include quantum aspects of gravity, semiclassical gravity remains the most fundamental description of nature currently available.

In this work, we adopt a semiclassical perspective. We assume that if a modified theory of gravity improves upon GR, its semiclassical extension should provide an even closer approximation to a fundamental description of nature. It is therefore essential to investigate whether modified gravity theories can be axiomatized and whether the methods of Hadamard renormalization can be extended to them. As a first step in this direction, we address these questions within Einstein--Cartan theory.

We organize the paper as follows. In Sec.~\ref{Preliminaries}, we provide a brief overview of semiclassical gravity and introduce some technical aspects. Section~\ref{Einstein--Cartan theory} presents the Einstein--Cartan theory, and in Sec.~\ref{Hadamard states}, we derive the corresponding Hadamard bi-parametrix. In Sec.~\ref{Core}, which constitutes the core of the paper, we introduce the axiomatic framework employed in the renormalization scheme, identify the resulting ambiguities, and review the conformal anomaly. Finally, Sec.~\ref{Conc} contains our conclusions. Some useful expressions are collected in Appendix~\ref{Appendix}.

Throughout this work, we use units where \( c = \hbar = G = 1 \), except in Subsec.~\ref{Scale and renormalization ambiguities} where $G\neq 1$. Abstract spacetime indices are denoted by lowercase Latin letters from the beginning of the alphabet. We consider a four-dimensional spacetime and use the metric \( g_{ab} \) and its inverse \( g^{ab} \) to lower and raise indices, respectively. We adopt the metric signature \( (-+++) \) and follow the curvature conventions of Ref.~\cite{Wald1984}. Symmetrization (antisymmetrization) is indicated by enclosing indices in parentheses (brackets), with a factor of \( 1/n! \), where \( n \) is the number of indices involved, excluding those between vertical bars. Finally, we assume that all fields are smooth and that spacetime is globally hyperbolic.

\section{Preliminaries} \label{Preliminaries}

In semiclassical GR, the equations of motion are~\cite{Wald1994}  
\begin{equation}
    \label{SEE}
    \mathring{G}_{ab} = 8 \pi \omega(\tau_{ab}),
\end{equation}  
where the left-hand side denotes the Einstein tensor constructed purely from the metric (we use the ring throughout the text to indicate quantities that only depend on the metric), and the right-hand side represents the expectation value, in a given state, of the operator associated with the energy--momentum tensor. In this way, both sides of Eq.~\eqref{SEE} are tensorial fields.

We begin by discussing the left-hand side of Eq.~\eqref{SEE}. The metric curvature tensor, \( {\mathring{R}_{abc}}^{\phantom{abc} d} \), is constructed using the torsion-free and metric-compatible derivative operator \( \mathring{\nabla}_a \). These two properties mean, respectively, that
\begin{equation}
(\mathring{\nabla}_a \mathring{\nabla}_b -\mathring{\nabla}_b \mathring{\nabla}_a) f  = 0,
\end{equation}
for any scalar function \( f \), and
\begin{equation}
\mathring{\nabla}_c g_{ab} = 0.
\end{equation}
The only nontrivial trace of \( {\mathring{R}_{abc}}^{\phantom{abc} d} \) is \( \mathring{R}_{ab} = {\mathring{R}_{acb}}^{\phantom{abc} c} \), which defines the Ricci tensor and satisfies \( \mathring{R}_{ab} = \mathring{R}_{ba} \). Moreover, the trace of the Ricci tensor is the Ricci scalar, $\mathring{R} = g^{ab} \mathring{R}_{ab}$. Finally, the Einstein tensor is given by
\begin{equation}
\mathring{G}_{ab} = \mathring{R}_{ab} - \frac{1}{2} \mathring{R} g_{ab}.
\end{equation}

Regarding the right-hand side of Eq.~\eqref{SEE}, there are some technical subtleties. In the particular case of a free Klein--Gordon field \( \Phi \), which we consider for simplicity, its associated energy--momentum tensor is quadratic in \( \Phi \)~\cite{Wald1984}. However, upon quantization this expression becomes ill-defined, since the corresponding ``quantum object,'' \( \hat{\Phi} \), is an operator-valued distribution. In other words, \( \hat{\Phi} \) only becomes an operator after acting on a test function. Hence, an expression such as
\begin{equation}
\hat{\Phi}^2[f] = \hat{\Phi}[\hat{\Phi}[f]],
\end{equation}
where \( f \) is a test function, is not well-defined. Consequently, no natural energy--momentum operator can be associated with the quantized free Klein--Gordon field.

As we shall see, this issue is resolved by acting with the energy--momentum operator on the Green function, \( G(x,x') \), and then taking the coincidence limit \( x' \rightarrow x \). This evaluation is divergent, and a renormalization procedure is required. To this end, Hadamard states are employed; these are states whose singular structure mirrors that of quantum fields in flat spacetime~\cite{Brown_1986,Decanini_2006,Decanini_2008}. This allows for the subtraction of the singular part (an alternative approach is discussed in Ref.~\cite{Johas}). Importantly, in semiclassical GR, an axiomatic framework exists~\cite{Wald1994} that leads to a well-defined expectation value for the energy--momentum tensor.

Our goal is to study semiclassical gravity in Einstein--Cartan theory, where spacetime torsion is nontrivial. The procedure we employ closely follows that of D\'ecanini and Folacci~\cite{Decanini_2008}. The most notable difference compared with GR is that the energy--momentum tensor is no longer divergence-free, a property that is repeatedly used in Ref.~\cite{Decanini_2008}. Instead, in Einstein--Cartan theory, its divergence satisfies a specific identity that also involves torsion and the spin--density tensor. We now turn to study some classical aspects of this theory.

\section{Einstein--Cartan Theory} \label{Einstein--Cartan theory}

This section reviews the classical Einstein--Cartan theory with a free, massive, nonminimally coupled Klein--Gordon field \( \Phi \). We begin by presenting some kinematical aspects.

\subsection{Kinematics}

We consider a nontrivial torsion tensor \( T^c_{\phantom{c} ab} \), defined by
\begin{equation}
    \label{Torsion}
   (\nabla_a \nabla_b - \nabla_b \nabla_a)f = -T^c_{\ ab} \nabla_c f,
\end{equation}
for any function \( f \). By definition, \( T^c_{\phantom{c} ab} = T^c_{\phantom{c} [ab]} \). Here, \( \nabla_a \) denotes a metric-compatible derivative operator with torsion. The operators \( \mathring{\nabla}_a \) and \( \nabla_a \) are related by
\begin{equation}
    \label{Connection3}
    (\nabla_a - \mathring{\nabla}_a)\, v_b = - K^{c}_{\phantom{c} ab}\, v_c ,
\end{equation}
for an arbitrary one-form \( v_a \), where
\begin{equation}
    K^{c}_{\phantom{c} ab}
    = \frac{1}{2} \left(T^{c}_{\phantom{c} ab} + T_{a \phantom{c} b}^{\phantom{a} c} + T_{b \phantom{c} a}^{\phantom{b} c}
      \right),
\end{equation}
is the contorsion tensor. Note that \( K_{cab} = K_{[c|a|b]} \), with \( K_{cab} = g_{cd} K^d_{\phantom{d} ab} \), and \( T^c_{\phantom{c} ab} = 2K^c_{\phantom{c} [ab]} \). Hence, torsion can always be determined from contorsion, and vice versa. Accordingly, in what follows, we use these two fields interchangeably.

The Riemann tensor associated with \( \nabla_a \) is defined by
\begin{equation}
    \label{Riemann1}
    R_{abc}^{\phantom{abc} d} \, v_d = \big( \nabla_a \nabla_b - \nabla_b \nabla_a + T^d_{\phantom{d} ab} \nabla_d \big) v_c,
\end{equation}
for any \( v_a \). Using Eq.~\eqref{Connection3}, we obtain
\begin{equation}
    \label{Riemann2}
    R_{abc}^{\phantom{abc} d} = \mathring{R}_{abc}^{\phantom{abc} d} - 2 \, \mathring{\nabla}_{[a} K^d_{\phantom{d} b] c} + 2 \, K^e_{\phantom{e} [a| c} K^d_{\phantom{d} |b] e}.
\end{equation}

In addition, we can readily verify that \( R_{abc}^{\phantom{abc} d} = R_{[ab] c}^{\phantom{[ab] c} d} \), and, from metric compatibility, \( R_{abcd} = R_{ab[cd]} \) (with \( R_{abcd} = g_{de} R_{abc}^{\phantom{abc} e} \)). Moreover,
\begin{align}
    R_{[abc]}^{\phantom{[abc]} d} &= \nabla_{[b} T^d_{\phantom{d} a]c} + T^e_{\phantom{e} [ab} T^d_{\phantom{d} c] e}, \\
    \nabla_{[a} R_{bc] d}^{\phantom{bc] d} e} &= T^f_{\phantom{f} [ab} R_{c] f d}^{\phantom{c] fd} e},
\end{align}
where the last equation is the corresponding Bianchi identity. Hence, \( R_{abcd} \neq R_{cdab} \), and thus \( R_{ab} \neq R_{(ab)} \). Finally, the Ricci tensor and Ricci scalar can be written as
\begin{align}
    R_{ab} &= \mathring{R}_{ab} - 2 \, \mathring{\nabla}_{[a} K^c_{\phantom{c} c] b} + 2 \, K^d_{\phantom{d} [a|b} K^c_{\phantom{c} |c] d}, \label{Ricci} \\
    R &= \mathring{R} + 2 \, \mathring{\nabla}_a K^{a \phantom{b} b}_{\phantom{a} b} + 2 \, K^{c \phantom{[a} a}_{\phantom{c} [a} K^b_{\phantom{b} b] c}. \label{ERicci}
\end{align}
We now turn to study the dynamics of the theory.

\subsection{Classical Dynamics} \label{EC-Dynamics}

The action we consider is
\begin{equation}
    \label{EHA}
    S[g^{ab}, K^c_{\phantom{c} ab}, \Phi] = \frac{1}{16 \pi} \int R \, \sqrt{-g} \, \mathrm{d}^4 x + S_{\text{M}},
\end{equation}
where the first term has the same form as the Einstein--Hilbert action of GR, but with a torsionful Ricci scalar, and 
\( S_{\text{M}} = \int \mathcal{L}_{\text{M}} \sqrt{-g}\,\mathrm{d}^4x \), 
with \( \mathcal{L}_{\text{M}} = \mathcal{L}_{\text{M}}[g^{ab},K^c_{\phantom{c} ab},\Phi] \). We take the variation of the action, neglecting boundary terms (as we do throughout the text), and obtain the equations of motion:
\begin{align}
    \mathring{G}_{ab} + K^d_{\phantom{d} (ab)} K^c_{\phantom{c} cd} - K^d_{\phantom{d} c(a} K^c_{\phantom{c} b)d} &- K^{d \phantom{[c} c}_{\phantom{d} [c} K^e_{\phantom{e} e]d} g_{ab} \nonumber \\
    &= 8 \pi \tau_{ab}, \label{Mg} \\
    g^{ab} K^d_{\phantom{d} dc} - \delta^a_{\phantom{a} c} K^{d \phantom{d} b}_{\phantom{d} d} + K_c^{\phantom{c} ba} - K^{b \phantom{c} a}_{\phantom{b} c} &= 16 \pi \sigma_c^{\phantom{c} ab}, \label{MK} \\
    \frac{\delta \mathcal{L}_{\text{M}}}{\delta \Phi} &= 0, \label{Mphi}
\end{align}
where the energy--momentum and spin--density tensors are defined, respectively, as
\begin{align}
    \label{Tau}
    \tau_{ab} &= -\frac{2}{\sqrt{-g}} \frac{\delta (\mathcal{L}_{\text{M}} \sqrt{-g})}{\delta g^{ab}},\\
    \sigma_c^{\phantom{c} ab} &= -\frac{1}{\sqrt{-g}} \frac{\delta (\mathcal{L}_{\text{M}} \sqrt{-g})}{\delta K^c_{\phantom{c} ab}}. \label{sigma}
\end{align}
We note that \( \tau_{ab} = \tau_{(ab)} \) and \( \sigma^{cab} = \sigma^{[c|a|b]} \), where \( \sigma^{cab} = g^{cd} \sigma_d^{\phantom{d} ab} \).

Equation~\eqref{Mg} generalizes Einstein's field equations, while Eq.~\eqref{Mphi} provides the equations of motion for the matter fields. Notably, Eq.~\eqref{MK} is algebraic, implying that the contorsion does not propagate (theories allowing for propagating torsion have been proposed; see, e.g., Ref.~\cite{NoPropTor}). In fact, we can invert Eq.~\eqref{MK} to show that \( K^c_{\phantom{c} ab} \neq 0 \) only where \( \sigma_c^{\phantom{c} ab} \neq 0 \), which typically occurs inside matter.

The matter action we consider is that of a free, massive Klein--Gordon field nonminimally coupled to (torsionful) curvature. Concretely,
\begin{equation}
	\label{SKG}
	S_{\text{KG}}  = -\frac{1}{2} \int \left( g^{ab} \mathring{\nabla}_a \Phi \mathring{\nabla}_b \Phi + (m^2 + \xi R) \Phi^2 \right) \sqrt{-g} \text{d}^4 x,
\end{equation}
where \( m \) is the field's mass and \( \xi \) is a dimensionless coupling constant. Thus, the scalar field equation of motion, known as the nonminimally coupled Klein--Gordon equation, is
\begin{equation}
    \label{KG}
    (\square - (m^2 + \xi R)) \Phi = 0,
\end{equation}
where \( \square = \mathring{\nabla}^a \mathring{\nabla}_a \). We can use Eq.~\eqref{ERicci} to write
\begin{align}
	\label{SKG2}
	S_{\text{KG}} &= -\frac{1}{2} \int \{g^{ab} \mathring{\nabla}_a \Phi \mathring{\nabla}_b \Phi + (m^2 + \xi \mathring{R}) \Phi^2 \nonumber\\
	&+ \xi g^{ab} [2(K^c_{\phantom{c} cb} \Phi \mathring{\nabla}_a \Phi - K^c_{\phantom{c} ab} \Phi \mathring{\nabla}_c \Phi) \nonumber\\
	&+ (K^d_{\phantom{d} ab} K^c_{\phantom{c} cd} - K^d_{\phantom{d} cb} K^c_{\phantom{c} ad}) \Phi^2] \} \sqrt{-g} \text{d}^4 x.
\end{align}
Hence, 
\begin{align}
    \label{TKG}
	\tau_{ab} &= \ (1 - 2 \xi) \mathring{\nabla}_a \Phi \mathring{\nabla}_b \Phi + \Bigg(2 \xi - \frac{1}{2} \Bigg) \mathring{\nabla}^c \Phi \mathring{\nabla}_c \Phi g_{ab}\nonumber\\
	&- 2 \xi \Phi \mathring{\nabla}_a \mathring{\nabla}_b \Phi + 2 \xi \Phi \square \Phi g_{ab} + \xi \Phi^2 G_{ab} - \frac{1}{2} m^2 \Phi^2 g_{ab} \nonumber\\
	&+ \xi [2(K^c_{\phantom{c} c(a} \Phi \mathring{\nabla}_{b)} \Phi - K^c_{\phantom{c} (ab)} \Phi \mathring{\nabla}_c \Phi) \nonumber\\
	&+ (K^d_{\phantom{d} (ab)} K^c_{\phantom{c} cd} - K^d_{\phantom{d} c(a} K^c_{\phantom{c} b)d}) \Phi^2 \nonumber\\
	&- (K^{d \phantom{[c} c}_{\phantom{d} [c} K^e_{\phantom{e} e]d} \Phi^2 - 2K^{c \phantom{d} d}_{\phantom{c} d} \Phi \mathring{\nabla}_c \Phi) g_{ab}],
\end{align}
and
\begin{align}
	\label{SigmaKG}
	\sigma_c^{\phantom{c} ab} &= \xi (\delta^a_{\phantom{a} c} \Phi \mathring{\nabla}^b \Phi - g^{ab} \Phi \mathring{\nabla}_c \Phi) \nonumber\\
	& + \frac{1}{2} \xi (g^{ab} K^d_{\phantom{d} dc} - \delta^a_{\phantom{a} c} K^{d \phantom{d} b}_{\phantom{d} d} + K_c^{\phantom{c} ba} - K^{b \phantom{c} a}_{\phantom{b} c}) \Phi^2.
\end{align}
Importantly, we take $m\geq 0$ and $\xi \neq 0$ so that torsion effects do not trivialize [see Eq.~\eqref{SigmaKG}].

Another point worth noting concerns the case of vanishing contorsion. When \( K^c_{\phantom{c} ab} = 0 \), Eq.~\eqref{SigmaKG} reduces to
\begin{equation}
    0 = \xi \big(\delta^a_{\phantom{a} c} \Phi \mathring{\nabla}^b \Phi - g^{ab} \Phi \mathring{\nabla}_c \Phi \big),
\end{equation}
which implies that the scalar field must be constant. Hence, contrary to naive expectations, setting \( K^c_{\phantom{c} ab} = 0 \) does not reproduce GR with a generic Klein--Gordon field. To obtain this theory, the contorsion must be set to zero before varying the action. This subtlety is important in the semiclassical analysis, because, within the axiomatic framework, we work directly with the equations of motion; thus, \( K^c_{\phantom{c} ab} = 0 \) does not reduce to the semiclassical Klein--Gordon field propagating in a torsionless spacetime.


\subsection{Symmetries}
\subsubsection{Invariance under diffeomorphisms}

Let us now turn our attention to the symmetries of the matter action, in particular its invariance under diffeomorphisms. This symmetry follows from the fact that the theory does not contain nondynamical fields~\cite{Lee,Cristobal}.

A generic matter action variation is
\begin{equation}
	\label{deltaSM}
	\delta S_{\text{M}} = -\int \Bigg\{\frac{1}{2} \tau_{ab} \delta g^{ab} + \sigma_c^{\phantom{c} ab} \delta K^c_{\phantom{c} ab} - \frac{\delta \mathcal{L}_{\text{M}}}{\delta \Phi} \delta \Phi \Bigg\} \sqrt{-g} \text{d}^4 x,
\end{equation}
where the last term vanishes on shell. For an infinitesimal diffeomorphism associated with (minus) the vector field \( \epsilon^{a} \), the dynamical fields transform with their Lie derivatives, namely,
    \begin{align}
	   \delta g^{ab} &=  2 \mathring{\nabla}^{(a} \epsilon^{b)}, \label{Lg}\\
	   \delta K^c_{\phantom{c} ab} &= -\epsilon^d \mathring{\nabla}_d K^c_{\phantom{c} ab} + K^d_{\phantom{d} ab} \mathring{\nabla}_d \epsilon^c\nonumber\\
        &- K^c_{\phantom{c} db} \mathring{\nabla}_a \epsilon^d - K^c_{\phantom{c} ad} \mathring{\nabla}_b \epsilon^d, \label{LK}\\
	   \delta \Phi &= -\epsilon^a \mathring{\nabla}_a \Phi, \label{Lphi}
    \end{align}
Inserting these results into Eq.~\eqref{deltaSM} and integrating by parts produces
\begin{align}
	\delta S_{\text{M}} &= -\int \epsilon^b \{-\mathring{\nabla}^a \tau_{ab} - \sigma_c^{\phantom{c} ad} \mathring{\nabla}_b K^c_{\phantom{c} ad} - \mathring{\nabla}_d K^d_{\phantom{d} ac} \sigma_b^{\phantom{b} ac}\nonumber\\
    &- K^d_{\phantom{d} ac} \mathring{\nabla}_d \sigma_b^{\phantom{b} ac} 
	+ \mathring{\nabla}_a K^c_{\phantom{c} bd} \sigma_c^{\phantom{c} ad} + K^c_{\phantom{c} bd} \mathring{\nabla}_a \sigma_c^{\phantom{c} ad}\nonumber\\
    &+ \mathring{\nabla}_d K^c_{\phantom{c} ab} \sigma_c^{\phantom{c} ad} + K^c_{\phantom{c} ab} \mathring{\nabla}_d \sigma_c^{\phantom{c} ad} \} \sqrt{-g} \text{d}^4 x.\label{IBD}
\end{align}
Thus, if \( S_{\text{M}} \) is invariant under an arbitrary diffeomorphism, then
\begin{align}
	\mathring{\nabla}^a \tau_{ab} &=  (K^c_{\phantom{c} bd} \mathring{\nabla}_a + K^c_{\phantom{c} ab} \mathring{\nabla}_d + \mathring{\nabla}_a K^c_{\phantom{c} bd} + \mathring{\nabla}_d K^c_{\phantom{c} ab}) \sigma_c^{\phantom{c} ad} \nonumber\\
    &- (K^d_{\phantom{d} ac} \mathring{\nabla}_d + \mathring{\nabla}_d K^d_{\phantom{d} ac}) \sigma_b^{\phantom{b} ac} - \mathring{\nabla}_b K^c_{\phantom{c} ad} \sigma_c^{\phantom{c} ad} \label{DivT}.
\end{align}
This is the identity that the energy–momentum and spin-density tensors must satisfy. Notably, in contrast with GR, \( \tau_{ab} \) is no longer necessarily divergence-free.

\subsubsection{Conformal transformations}

We now consider conformal transformations of the matter action. Under these transformations, the fields change as
\begin{equation*}
    g^{ab} \rightarrow \Omega^{-2} g^{ab}, \quad K^c_{\phantom{c} ab} \rightarrow K^c_{\phantom{c} ab}, \quad \Phi \rightarrow \Omega^{-1} \Phi,
\end{equation*}
where $\Omega$ is a strictly positive function and the conformal weight on the scalar field is set so that the Klein--Gordon equation is also conformally invariant \cite[Appendix D]{Wald1984}. For infinitesimal conformal transformations, we take $\Omega = 1 + \epsilon + \mathcal{O}(\epsilon^2)$, for any infinitesimal function $\epsilon$, which gives
    \begin{align}
        \delta g^{ab} &=- 2 \epsilon g^{ab}, \label{TCg}\\
        \delta K^c_{\phantom{c} ab} &= 0, \label{TCK}\\
        \delta \Phi &= -\epsilon \Phi. \label{TCphi}
    \end{align}

In this case, the on-shell variation of the matter action, Eq.~\eqref{deltaSM}, becomes
\begin{equation}
    \label{TCSM}
    \delta S_{\text{M}} = \int \epsilon \, \tau^a_{\phantom{a} a} \, \sqrt{-g} \, \mathrm{d}^4 x.
\end{equation}
Hence, for \( S_{\text{M}} \) to be invariant under an arbitrary conformal transformation, we require
\begin{equation}
    \label{Trtau}
    \tau^a_{\phantom{a} a} = 0.
\end{equation}

For the Klein--Gordon field under consideration in the special case where \( m = 0 \) and \( \xi = 1/6 \), and the fields are on-shell, this reduces to
\begin{equation}
    \label{TrTKGC1}
    \tau^a_{\phantom{a} a} = \frac{1}{3} \mathring{\nabla}_a \big( K^{a \phantom{b} b}_{\phantom{a} b} \, \Phi^2 \big),
\end{equation}
which is a total divergence. Since we neglect such boundary terms, we can conclude that, for \( m = 0 \) and \( \xi = 1/6 \), the classical theory is conformally invariant.

Thus far, we have worked with a classical theory for both the gravitational and matter fields. We now proceed to consider the quantum aspects of the matter fields by introducing the Hadamard bi-parametrix.


\section{Hadamard bi-parametrix} \label{Hadamard states}

The formalism of Hadamard renormalization is most naturally formulated within the algebraic approach to QFTCS~\cite{Haag1}, which bypasses the need to introduce a particular notion of positive frequency when constructing the associated Hilbert space. We work with the Klein--Gordon algebra, \( \mathcal{A}_{\text{KG}} \), generated by the operators \( \hat{\Phi}[\cdot] \), which act on complex test functions of compact support and satisfy:
\begin{enumerate}
    \item \( f \mapsto \hat{\Phi}[f] \) is linear,
    \item \( \hat{\Phi}[f]^* = \hat{\Phi}[\bar{f}] \), where the star denotes the algebra adjoint and the bar denotes complex conjugation,
    \item \( \hat{\Phi}\left[(\square - (m^2 + \xi R))f\right] = 0 \), i.e., the Klein--Gordon equation is satisfied,
    \item \( [\hat{\Phi}[f], \hat{\Phi}[g]] = -i \, \text{E}\left([f, g]\right)  \) for \(f, g \) functions of compact support, where \( \text{E}^\pm \) are the retarded (\(-\)) and advanced (\(+\)) Green operators of Eq.~\eqref{KG}, and \( \text{E} = \text{E}^+ - \text{E}^- \) is the causal propagator.
\end{enumerate}

To obtain well-defined operators associated with \( \tau_{ab} \) and \( \sigma_c^{\phantom{c} ab} \), which are quadratic in \( \hat{\Phi} \), we employ point-splitting renormalization \cite{Christensen_1976}. In this scheme, the central object is the two-point function
\begin{align}
    \label{FDP}
    \omega (\hat{\Phi}[f_1] \hat{\Phi}[f_2]) =& \int G(x, x') f_1(x) f_2(x') \nonumber\\
    &\times \sqrt{-g(x)} \sqrt{-g(x')} \, \mathrm{d}^4 x \, \mathrm{d}^4 x',
\end{align}
where \( \omega \) denotes the expectation value in a given state, and both, $f_1$ and $f_2$, are ``smearing functions'' of compact support. Recall that \( G(x, x') \) is the Green function, which satisfies
\begin{align}
    \label{Green}
    (\square_x - (m^2 + \xi R(x))) G(x, x') &= -\delta^4(x, x'), \nonumber\\
    (\square_{x'} - (m^2 + \xi R(x'))) G(x, x') &= -\delta^4(x, x'),
\end{align}
with \( \delta^4(x, x') = - \delta^4(x - x') / \sqrt{-g(x)} \). We adopt the prescription in which the Green function coincides with the Feynman propagator, denoted by \( G^{\text{F}}(x, x') \). This Green function satisfies \( G^{\text{F}}(x, x') = G^{\text{F}}(x', x) \). Moreover, \( G^{\text{F}}(x, x') \to \infty \) as \( x' \to x \); this is known as the coincidence limit.

Hadamard states for the Klein--Gordon field are defined so that, in the coincidence limit, the singular structure of \( G^{\text{F}}(x, x') \) coincides with that in flat spacetime. This definition is justified by the local flatness of spacetime. The singular structure is given by
\begin{align}
    \label{Hadamard}
    G^{\text{F}}(x, x') 
    & \to \frac{i}{8 \pi^2} \Bigg( \frac{U(x, x')}{\sigma(x, x') + i \varepsilon} 
    \nonumber\\
    &+ V(x, x') \ln\Big(\frac{\sigma(x, x')}{\ell^2} + i \varepsilon\Big) 
    + W(x, x') \Bigg).
\end{align}
where \( \sigma(x, x') \) is half the squared geodesic distance between \( x \) and \( x' \), and \( \ell > 0 \) is an arbitrary length scale introduced to render the argument of the logarithm dimensionless. Also, \( i \varepsilon \) is introduced so that $G^{\text{F}}(x, x')$ is consistent with the Feynman prescription as \( \varepsilon \to 0^+ \)~\cite{fulling}. Moreover, \( U(x, x') \), \( V(x, x') \), and \( W(x, x') \) are smooth, regular in the coincidence limit, and symmetric under \( x \leftrightarrow x' \).

The Hadamard bi-parametrix associated with the length scale $\ell$ is defined as the divergent part of Eq.~\eqref{Hadamard}, namely,
\begin{align}
    \label{HB}
    H_\ell(x, x') &= \frac{i}{8 \pi^2} \Bigg( \frac{U(x, x')}{\sigma(x, x') + i \varepsilon} 
    \nonumber\\
   &+ V(x, x') \ln\Big(\frac{\sigma(x, x')}{\ell^2} + i \varepsilon\Big)  \Bigg).
\end{align}
Note that, under a change of ``scale'', i.e., shifting the length scale in the logarithm, we get
\begin{equation}
    \label{ScaleAmbiguities}
    H_\ell(x, x') = H_{\ell'}(x, x') + \frac{i}{8 \pi^2} V(x, x') \ln(M^2),
\end{equation}
where \( M = \ell' / \ell \).

We now insert Eq.~\eqref{Hadamard} into Eq.~\eqref{Green} for \( x \neq x' \). From Eqs.~\eqref{PropSig1} and~\eqref{PropSig2} in Appendix~\ref{Appendix}, we obtain  
\begin{align}
    0 &= \frac{-2 \mathring{\nabla}_a U \, \mathring{\nabla}^a \sigma + 2U \, \Delta^{-1/2} \mathring{\nabla}_a \Delta^{1/2} \, \mathring{\nabla}^a \sigma}{\sigma} \nonumber\\
    &\quad + (\square - (m^2 + \xi R)) V \, \sigma \ln (\sigma) \nonumber\\
    &\quad + \sigma (\square - (m^2 + \xi R)) W + (\square - (m^2 + \xi R)) U \nonumber\\
    &\quad + 2 \mathring{\nabla}_a V \, \mathring{\nabla}^a \sigma + 2V \left[1 - \Delta^{-1/2} \mathring{\nabla}_a \Delta^{1/2} \, \mathring{\nabla}^a \sigma \right],
\end{align}
where \( \Delta \) is the Van~Vleck--Morette determinant (see Appendix~\ref{Appendix}). All arguments are omitted for brevity. By collecting the terms with the same $\sigma$-dependence, we can extract the following relations:
\begin{align}
    U &= \Delta^{1/2}, \label{U}\\
    (\square - (m^2 + \xi R)) V &= 0, \label{V}\\
    \sigma (\square - (m^2 + \xi R)) W &= - (\square - (m^2 + \xi R)) U \nonumber\\
     - 2 \Big[ \mathring{\nabla}_a V \, \mathring{\nabla}^a \sigma & - V (1 - \Delta^{-1/2} \mathring{\nabla}_a \Delta^{1/2} \, \mathring{\nabla}^a \sigma) \Big]. \label{W}
\end{align}

To analyze the structure of \( V \) and \( W \), we expand them as
\begin{align}
    V(x, x') &= \sum_{n=0}^{\infty} V_n(x, x') \, \sigma(x, x')^n, \label{VSeries}\\
    W(x, x') &= \sum_{n=0}^{\infty} W_n(x, x') \, \sigma(x, x')^n. \label{WSeries}
\end{align}
The recursion relations for the expansion coefficients follow from inserting Eqs.~\eqref{VSeries} and~\eqref{WSeries} into Eq.~\eqref{Green}. This procedure yields
\begin{align}
    0 &= 2(n+1)(n+2) V_{n+1} + 2(n+1) \mathring{\nabla}_a V_{n+1} \, \mathring{\nabla}^a \sigma \nonumber\\
    &- 2(n+1) V_{n+1} \Delta^{-1/2} \mathring{\nabla}_a \Delta^{1/2} \, \mathring{\nabla}^a \sigma\nonumber\\
    &+ (\square - (m^2 + \xi R)) V_n, \label{Vn} \\
    0 &= 2(n+1)(n+2) W_{n+1} + 2(n+1) \mathring{\nabla}_a W_{n+1} \, \mathring{\nabla}^a \sigma \nonumber\\
    &- 2(n+1) W_{n+1} \Delta^{-1/2} \mathring{\nabla}_a \Delta^{1/2} \, \mathring{\nabla}^a \sigma + 2(2n+3) V_{n+1} \nonumber\\
    &+ 2 \mathring{\nabla}_a V_{n+1} \, \mathring{\nabla}^a \sigma - 2 V_{n+1} \Delta^{-1/2} \mathring{\nabla}_a \Delta^{1/2} \, \mathring{\nabla}^a \sigma \nonumber\\
    &+ (\square - (m^2 + \xi R)) W_n, \label{Wn}
\end{align}
subject to the condition
\begin{align}
    \label{BC}
    0 &= 2 \Big[ \mathring{\nabla}_a V_0 \, \mathring{\nabla}^a \sigma + V_0 (1 - \Delta^{-1/2} \mathring{\nabla}_a \Delta^{1/2} \, \mathring{\nabla}^a \sigma) \Big] \nonumber\\
    &+ (\square - (m^2 + \xi R)) U.
\end{align}
Note that both, \( U \) and \( V \), are geometric and state-independent: \( U \) is the square root of the Van~Vleck--Morette, and we can obtain \( V \) from the recursion relations in Eq.~\eqref{Vn} together with the ``boundary'' condition in Eq.~\eqref{BC}. We emphasize that \( V \) depends on \( m \) and \( \xi \).

The function \( W \) is the only one that depends on the state. In fact, \( W_0 \) is the sole independent expansion coefficient; the remaining terms in \( W \) are determined from \( W_0 \) via Eq.~\eqref{Wn}. Moreover, \( W \) satisfies an identity obtained by extracting \( (\square - (m^2 + \xi R)) U \) from Eq.~\eqref{BC}, substituting this into Eq.~\eqref{W}, and then expanding \( V \) using Eq.~\eqref{VSeries}. The resulting identity is
\begin{equation}
    \label{WV1}
    (\square - (m^2 + \xi R))W = -6V_1 - 2 \mathring{\nabla}_a V_1 \, \mathring{\nabla}^a \sigma + \mathcal{O} (\sigma).
\end{equation}
A similar identity holds for \( V_0 \), which can be derived by inserting Eq.~\eqref{VSeries} into Eq.~\eqref{V}, yielding
\begin{equation}
    \label{VV1}
    (\square - (m^2 + \xi R))V_0 = -4V_1 - 2 \mathring{\nabla}_a V_1 \, \mathring{\nabla}^a \sigma + \mathcal{O} (\sigma).
\end{equation}
We retain these identities with terms up to \( \mathcal{O} (\sigma^{1/2}) \), since terms of higher order in \( \sigma \) vanish in the coincidence limit.

The coefficients in Eqs.~\eqref{VSeries} and~\eqref{WSeries} can be further expanded in a covariant Taylor series:
\begin{align}
    V_n(x, x') &= v_n(x) + \sum_{p=1}^{\infty} \frac{(-1)^p}{p!} {v_n}_{a_1 \cdots a_p}(x) \, 
    \mathring{\nabla}^{a_1} \sigma(x, x') \cdots \mathring{\nabla}^{a_p} \sigma(x, x'), \label{vnp} \\
    W_n(x, x') &= w_n(x) + \sum_{p=1}^{\infty} \frac{(-1)^p}{p!} {w_n}_{a_1 \cdots a_p}(x) \, 
    \mathring{\nabla}^{a_1} \sigma(x, x') \cdots \mathring{\nabla}^{a_p} \sigma(x, x'). \label{wnp}
\end{align}
Using Eqs.~\eqref{vnp} and \eqref{PropSig3}, the right-hand sides of Eqs.~\eqref{WV1} and \eqref{VV1} become
\begin{align}
    \label{Wv1}
    (\square - (m^2 + \xi R))W &= -6v_1 + 2 \mathring{\nabla}_a v_1 \, \mathring{\nabla}^a \sigma + \mathcal{O} (\sigma), \\
    \label{Vv1}
    (\square - (m^2 + \xi R))V_0 &= -4v_1 + \mathring{\nabla}_a v_1 \, \mathring{\nabla}^a \sigma + \mathcal{O} (\sigma).
\end{align}
Note that one of these equations is for a single term in the expansion, $V_0$, while the other is for the entire object, $W$.

From Eqs.~\eqref{vnp} and~\eqref{wnp}, we can verify that \( q_{a_1 \cdots a_p} = q_{(a_1 \cdots a_p)} \), where \( q \) stands for either \( v_n \) or \( w_n \). This symmetry, together with the exchange symmetry \( x \leftrightarrow x' \) of \( V_0 \) and \( W \), allows us to express the coefficients with an odd number of indices in terms of those with an even number of indices. This is achieved by using Eqs.~\eqref{PropSig3}-\eqref{PropSig10} given in Appendix \ref{Appendix}, explicitly implementing the exchange symmetry in Eqs.~\eqref{vnp} and~\eqref{wnp}, and then taking the coincidence limit. The resulting identities are
\begin{equation}
    \label{OddtoEven}
    q_a = \frac{1}{2} \mathring{\nabla}_a q, \quad 
    q_{abc} = \frac{3}{2} \mathring{\nabla}_{(c} q_{ab)} - \frac{1}{4} \mathring{\nabla}_{(a} \mathring{\nabla}_b \mathring{\nabla}_{c)} q.
\end{equation}
These relations are particularly useful when expanding the left-hand sides of Eqs.~\eqref{Wv1} and~\eqref{Vv1} up to \( p = 3 \). From these results, it follows in the coincidence limit that
    \begin{align}
        w^a_{\phantom{a} a} &= (m^2 + \xi \mathring{R}) w - 6v_1 
        + 2 \xi (\mathring{\nabla}_a K^{a \phantom{b} b}_{\phantom{a} b} + K^{c \phantom{[a} a}_{\phantom{c} [a} K^b_{\phantom{b} b] c}) w, \label{wtrace}\\
        \mathring{\nabla}_b w^b_{\phantom{b} a} &= \frac{1}{4} \mathring{\nabla}_a \square w + \frac{1}{2} \mathring{R}_{ab} \mathring{\nabla}^b w + \frac{1}{2} \xi \mathring{\nabla}_a \mathring{R} w - \mathring{\nabla}_a v_1  + \xi (\mathring{\nabla}_a \mathring{\nabla}_b K^{b \phantom{c} c}_{\phantom{b} c} + \mathring{\nabla}_a K^{d \phantom{[b} b}_{\phantom{d} [b} K^c_{\phantom{c} c] d} 
        + K^{d \phantom{[b|} b}_{\phantom{d} [b|} \mathring{\nabla}_a K^c_{\phantom{c} |c] d}) w, \label{wdiv}\\
        {v_0}^a_{\phantom{a} a} &= (m^2 + \xi \mathring{R}) v_0 - 4v_1 
        + 2 \xi (\mathring{\nabla}_a K^{a \phantom{b} b}_{\phantom{a} b} + K^{c \phantom{[a} a}_{\phantom{c} [a} K^b_{\phantom{b} b] c}) v_0, \label{vtrace}\\
        \mathring{\nabla}_b {v_0}^b_{\phantom{b} a} &= \frac{1}{4} \mathring{\nabla}_a \square v_0 + \frac{1}{2} \mathring{R}_{ab} \mathring{\nabla}^b v_0 + \frac{1}{2} \xi \mathring{\nabla}_a \mathring{R} v_0 - \mathring{\nabla}_a v_1+ \xi (\mathring{\nabla}_a \mathring{\nabla}_b K^{b \phantom{c} c}_{\phantom{b} c} + \mathring{\nabla}_a K^{d \phantom{[b} b}_{\phantom{d} [b} K^c_{\phantom{c} c] d} 
        + K^{d \phantom{[b|} b}_{\phantom{d} [b|} \mathring{\nabla}_a K^c_{\phantom{c} |c] d}) v_0. \label{vdiv}
    \end{align}

Finally, since \( U \) and \( V \) are geometric quantities, Eq.~\eqref{BC} can be used to determine \( v_0 \) and \( {v_{0}}_{ab} \), which in turn allows us to find \( v_1 \) using Eq.~\eqref{vtrace}. Expanding Eq.~\eqref{BC} and utilizing Eq.~\eqref{vnp} yields
\begin{equation}
    \label{BDT}
    2v_0 + \Big(3{v_{0}}_{ab} - \mathring{\nabla}_a \mathring{\nabla}_b v_0 - \frac{1}{3} \mathring{R}_{ab} v_0 \Big) \mathring{\nabla}^a \sigma \mathring{\nabla}^b \sigma  + (\square - (m^2 + \xi R)) \Delta^{1/2} = 0.
\end{equation}
Then, using Eqs.~\eqref{VMDE}-\eqref{DDSigE}, we obtain
    \begin{align}
        v_0 &= \frac{1}{2} (m^2 + \xi R) - \frac{1}{12} \mathring{R}, \label{v0}\\
        {v_{0}}_a &= \frac{1}{4} \xi \mathring{\nabla}_a R - \frac{1}{24} \mathring{\nabla}_a \mathring{R}, \label{v0a}\\
        {v_{0}}_{ab} &= \frac{1}{12} m^2 \mathring{R}_{ab} + \frac{1}{6} \xi \mathring{\nabla}_a \mathring{\nabla}_b R - \frac{1}{40} \mathring{\nabla}_a \mathring{\nabla}_b \mathring{R} - \frac{1}{120} \square \mathring{R}_{ab} + \frac{1}{12} \xi R \mathring{R}_{ab} - \frac{1}{72} \mathring{R} \mathring{R}_{ab} + \frac{1}{90} \mathring{R}_a^{\phantom{a} c} \mathring{R}_{bc}\nonumber\\
        &- \frac{1}{180} \mathring{R}^{cd} \mathring{R}_{acbd} - \frac{1}{180} \mathring{R}_a^{\phantom{a} cde} \mathring{R}_{bcde}, \label{v0ab}\\
        v_1 &= \frac{1}{8} m^4 + \frac{1}{4} \xi m^2 R - \frac{1}{24} m^2 \mathring{R} - \frac{1}{24} \xi \square R + \frac{1}{120} \square \mathring{R}  + \frac{1}{8} \Big(\xi R - \frac{1}{6} \mathring{R} \Big)^2 - \frac{1}{720} \mathring{R}_{ab} \mathring{R}^{ab} + \frac{1}{720} \mathring{R}_{abcd} \mathring{R}^{abcd}. \label{v1}
    \end{align}

With the results obtained in this section, we have the relevant expressions to perform the Hadamard renormalization of the expectation values of the energy--momentum and spin--density operators. In the following section, we define these operators and provide an axiomatic framework that ensures that their expectation values are well defined.
 
\section{Semiclassical Einstein--Cartan theory} \label{Core}

The semiclassical analysis we consider involves quantizing the Klein--Gordon field of the theory described in Sec.~\ref{Einstein--Cartan theory}. The semiclassical dynamics is governed by
\begin{align}
    \mathring{G}_{ab} 
    + K^d_{\phantom{d} (ab)} K^c_{\phantom{c} cd} 
    - K^d_{\phantom{d} c (a} K^c_{\phantom{c} b) d} 
    &- K^{d\ \ c}_{\ [c} K^e_{\phantom{e} e] d} g_{ab} 
    \nonumber\\
    &= 8\pi\, \omega(\tau_{ab}), \label{SMg}\\
    g^{ab} K^d_{\phantom{d} dc} 
    - \delta^a_{\phantom{a} c} K^{d\ b}_{\ d} 
    + K_c^{\phantom{c} ba} 
    - K^{b\ a}_{\ c} 
    &= 16\pi\, \omega(\sigma_c^{\phantom{c} ab}), \label{SMK}\\
    \hat{\Phi}\big[(\square - (m^2 + \xi R)) f\big] &= 0, \label{SMphi}
\end{align}
where the last equation holds for any function $f$ of compact support, and \( \omega \) denotes the (renormalized) expectation value with respect to a Hadamard state. We now present the axiomatic framework that leads to these expressions.

\subsection{Axiomatic framework} \label{Axiomatic framework}


Here, we generalize Wald's axioms~\cite{Wald1994} to a theory with a torsionful connection. We propose the following set of axioms:
\begin{enumerate}
    \item If the expectation values in two different Hadamard states, \( \omega_1 \) and \( \omega_2 \), are such that 
    \[
        \omega_1(\hat{\Phi}[f_1]\hat{\Phi}[f_2]) - \omega_2(\hat{\Phi}[f_1]\hat{\Phi}[f_2])
    \]
    is smooth for any pair of test functions of compact support, \( f_1\), \(f_2 \), then the differences 
    \(\omega_1(\tau_{ab}) - \omega_2(\tau_{ab})\) and 
    \(\omega_1(\sigma_c^{\phantom{c} ab}) - \omega_2(\sigma_c^{\phantom{c} ab})\) are also smooth.

    \item The expectation values \( \omega(\tau_{ab}(x)) \) and \( \omega(\sigma_c^{\phantom{c} ab}(x)) \) are local with respect to the state of the Klein--Gordon field in the following sense:  
    Let \( (\mathcal{M}, g_{ab}) \) and \( (\mathcal{M}', g_{ab}') \) be globally hyperbolic spacetimes with Cauchy surfaces \( \Sigma \) and \( \Sigma' \), respectively.  
    Let \( \mathcal{O} \subset \mathcal{M} \) and \( \mathcal{O}' \subset \mathcal{M}' \) be globally hyperbolic open neighborhoods of \( x \) and \( x' \), with Cauchy surfaces \( \mathcal{O} \cap \Sigma \) and \( \mathcal{O}' \cap \Sigma' \), such that there exists an isometry between \( \mathcal{O} \) and \( \mathcal{O}' \).  
    Under this isometry, we can identify the Klein--Gordon subalgebras 
    \({\mathcal{A}_{\text{KG}}}_{\mathcal{O}} \subset \mathcal{A}_{\text{KG}}\) for \( \mathcal{M} \) and 
    \({\mathcal{A}'_{\text{KG}}}_{\mathcal{O}'} \subset \mathcal{A}'_{\text{KG}}\) for \( \mathcal{M}' \).  
    If the restrictions of the Hadamard states coincide, i.e.
    \(\omega|_{{\mathcal{A}_{\text{KG}}}_{\mathcal{O}}} 
      = \omega'|_{{\mathcal{A}'_{\text{KG}}}_{\mathcal{O}'}}\),
    then we require that
    \(\omega(\tau_{ab}(x)) = \omega'(\tau_{ab}(x'))\) and 
    \(\omega(\sigma_c^{\phantom{c} ab}(x)) = \omega'(\sigma_c^{\phantom{c} ab}(x'))\).

    \item For any Hadamard state, the expectation values satisfy
    \begin{align}
        & \mathring{\nabla}^a \omega(\tau_{ab}) = \nonumber\\
       & (K^c_{\phantom{c} bd} \mathring{\nabla}_a 
        + K^c_{\phantom{c} ab} \mathring{\nabla}_d 
        + \mathring{\nabla}_a K^c_{\phantom{c} bd} 
        + \mathring{\nabla}_d K^c_{\phantom{c} ab})
        \, \omega(\sigma_c^{\phantom{c} ad}) \nonumber\\
        &- (K^d_{\phantom{d} ac} \mathring{\nabla}_d 
        + \mathring{\nabla}_d K^d_{\phantom{d} ac})
        \, \omega(\sigma_b^{\phantom{b} ac})
        - \mathring{\nabla}_b K^c_{\phantom{c} ad} \,
        \omega(\sigma_c^{\phantom{c} ad}).
    \end{align}

    \item For a vacuum \( \Omega \) in flat, torsionless spacetime,
    \(\omega_{\Omega}(\tau_{ab}) = 0\) and 
    \(\omega_{\Omega}(\sigma_c^{\phantom{c} ab}) = 0\).
\end{enumerate}

The first two axioms determine \( \omega(\tau_{ab}) \) and \( \omega(\sigma_c^{\phantom{c} ab}) \) up to local curvature terms, which cannot diverge in the flat-spacetime limit, as required by the fourth axiom. The third axiom is directly inspired by the (classical) relation given in Eq.~\eqref{DivT}. Moreover, the fourth axiom ensures the consistency of the semiclassical approach with quantum field theory in torsion-free flat spacetime. Equipped with these axioms, we now perform the regularization of the expectation values of the energy--momentum and spin--density operators.

\subsection{Renormalization}

To compute the expectation values of the energy--momentum and spin--density operators, we employ a point-splitting regularization procedure~\cite{Christensen_1976}. This procedure involves acting on \( G^{\text{F}}(x, x') \) with the following differential operators, which we obtain directly from the classical expressions in Eqs.~\eqref{TKG} and~\eqref{SigmaKG}:
\begin{align}
		\hat{\tau}_{ab} &= (1 - 2 \xi) g_b^{\phantom{b} b'} \mathring{\nabla}_a \mathring{\nabla}_{b'} + \Bigg(2 \xi - \frac{1}{2} \Bigg) g_{ab} g^{cd'} \mathring{\nabla}_c \mathring{\nabla}_{d'} \nonumber\\
    &- 2 \xi g_a^{\phantom{a} a'} g_b^{\phantom{b} b'} \mathring{\nabla}_{a'} \mathring{\nabla}_{b'} + 2 \xi g_{ab} \square + \xi \mathring{G}_{ab} - \frac{1}{2} m^2 g_{ab} \nonumber\\
	&+ \xi [K^c_{\phantom{c} c(a} \mathring{\nabla}_{b)} + K^c_{\phantom{c} c(a} g_{b)}^{\phantom{b)} a'} \mathring{\nabla}_{a'} \nonumber\\
    &- K^c_{\phantom{c} (ab)} \mathring{\nabla}_c - K^c_{\phantom{c} (ab)} g_c^{\phantom{c} c'} \mathring{\nabla}_{c'} \nonumber\\
    &+ K^d_{\phantom{d} (ab)} K^c_{\phantom{c} cd} - K^d_{\phantom{d} c (a} K^c_{\phantom{c} b) d} \nonumber\\
    &- g_{ab} (K^{d \phantom{[c} c}_{\phantom{d} [c} K^e_{\phantom{e} e] d} - K^{c \phantom{d} d}_{\phantom{c} d} \mathring{\nabla}_c - K^{c \phantom{d} d}_{\phantom{c} d} g_c^{\phantom{c} c'} \mathring{\nabla}_{c'})],
    \label{TauHat}\\
		\hat{\sigma}_c^{\phantom{c} ab} &= \frac{1}{2} \xi (\delta^a_{\phantom{a} c} \mathring{\nabla}^b + \delta^a_{\phantom{a} c} g_{b'}^{\phantom{b'} b} \mathring{\nabla}^{b'} - g^{ab} \mathring{\nabla}_c - g^{ab} g_c^{\phantom{c} c'} \mathring{\nabla}_{c'} \nonumber\\
    &+ g^{ab} K^d_{\phantom{d} dc} - \delta^a_{\phantom{a} c} K^{d \phantom{d} b}_{\phantom{d} d} + K_c^{\phantom{c} ba} - K^{b \phantom{c} a}_{\phantom{b} c}),\label{SigmaHat}
\end{align}
where unprimed indices refer to \( x \), and primed indices to \( x' \). In addition, \( g_{a}^{\ a'} \) is the bitensor that implements parallel transport from \( x \) to \( x' \), as described in Appendix~\ref{Appendix}. Concretely, the expectation values we need to compute are
\begin{align}
    \omega (\tau_{ab}) &= \lim_{x' \rightarrow x} \hat{\tau}_{ab} (x, x') [-iG^{\text{F}} (x, x')], \label{omegatau}\\
    \omega (\sigma_c^{\phantom{c} ab}) &= \lim_{x' \rightarrow x} \hat{\sigma}_c^{\phantom{c} ab} (x, x') [-iG^{\text{F}} (x, x')]. \label{omegasigma}
\end{align}

As we mention above, \( G^{\text{F}} (x, x') \) is singular in the coincidence limit; hence, the above expressions are also singular. However, we can define the regularized expectation values (represented with the same symbols, for simplicity) as
\begin{align}
    \omega (\tau_{ab}) &= \lim_{x' \rightarrow x} \hat{\tau}_{ab} (x, x') [-i(G^{\text{F}} (x, x') - H_\ell (x, x'))] \nonumber\\
    &= \frac{1}{8 \pi^2} \lim_{x' \rightarrow x} \hat{\tau}_{ab} (x, x') W(x, x') + \tilde{\Theta}_{ab}, \label{omegataureg}\\
    \omega (\sigma_c^{\phantom{c} ab}) &= \lim_{x' \rightarrow x} \hat{\sigma}_c^{\phantom{c} ab} (x, x') [-i(G^{\text{F}} (x, x') - H_\ell (x, x'))] \nonumber\\
    &= \frac{1}{8 \pi^2} \lim_{x' \rightarrow x} \hat{\sigma}_c^{\phantom{c} ab} (x, x') W(x, x') + \tilde{\Sigma}_c^{\phantom{c} ab}, \label{omegasigmareg}
\end{align}
where \( \tilde{\Theta}_{ab} \) and \( \tilde{\Sigma}_c^{\phantom{c} ab} \) are finite ambiguities that arise from the regularization procedure. We can decompose these ambiguities as
\begin{align}
    \tilde{\Theta}_{ab} &= \Theta^{M^2}_{ab} + \Theta_{ab} + \frac{1}{4 \pi^2} g_{ab} v_1, \label{ambTheta}\\
    \tilde{\Sigma}_c^{\phantom{c} ab} &= {\Sigma^{M^2}}_c^{\phantom{c} ab} + \Sigma_c^{\phantom{c} ab}, \label{ambSigma}
\end{align}
where \( \Theta^{M^2}_{ab} \) and \( {\Sigma^{M^2}}_c^{\phantom{c} ab} \) are scale ambiguities, introduced to account for Eq.~\eqref{ScaleAmbiguities}, while \( \Theta_{ab} \) and \( \sigma_c^{\phantom{c} ab} \) correspond to renormalization ambiguities. We explain the appearance of the term \( g_{ab} v_1/(4 \pi^2) \) below.

We can explicitly compute \( \omega (\tau_{ab}) \) and \( \omega (\sigma_c^{\phantom{c} ab}) \) by expanding \( W \) using Eq.~\eqref{wnp}. Additionally, using Eqs.~\eqref{PropSig3}-\eqref{PropSig7}, we can bring these expectation values to the form
\begin{align}
	\omega (\tau_{ab}) &= \frac{1}{8 \pi^2} \Bigg\{-\Bigg(w_{ab} - \frac{1}{2} g_{ab} w^c_{\phantom{c} c} \Bigg) \nonumber\\
    &+ \frac{1}{2} (1 - 2 \xi) \mathring{\nabla}_a \mathring{\nabla}_b w + \frac{1}{2} \Bigg(2 \xi - \frac{1}{2} \Bigg) g_{ab} \square w \nonumber\\
    & + \xi \mathring{G}_{ab} w - \frac{1}{2} m^2 g_{ab} w \nonumber\\
    &+ \xi [K^c_{\phantom{c} c (a} \mathring{\nabla}_{b)} w - K^c_{\phantom{c} (ab)} \mathring{\nabla}_c w \nonumber\\
    &+ (K^d_{\phantom{d} (ab)} K^c_{\phantom{c} cd} - K^d_{\phantom{d} c (a} K^c_{\phantom{c} b) d}) w \nonumber\\
	&- g_{ab} (K^{d \phantom{[c} c}_{\phantom{d} [c} K^e_{\phantom{e} e] d} w - K^{c \phantom{d} d}_{\phantom{c} d} \mathring{\nabla}_c w)] \Bigg\} + \tilde{\Theta}_{ab},\label{TEMECS}\\
	\omega (\sigma_c^{\phantom{c} ab}) &= \frac{1}{16 \pi^2} \xi \Big\{\delta^a_{\phantom{a} c} \mathring{\nabla}^b w - g^{ab} \mathring{\nabla}_c w \nonumber\\
    &+ (g^{ab} K^d_{\phantom{d} dc} - \delta^a_{\phantom{a} c} K^{d \phantom{d} b}_{\phantom{d} d} + K_c^{\phantom{c} ba} - K^{b \phantom{c} a}_{\phantom{b} c}) w \Big\} \nonumber\\
    & + \tilde{\Sigma}_c^{\phantom{c} ab}. \label{DSECS}
\end{align}

Furthermore, we can determine the condition required for the third axiom to be satisfied using Eqs.~\eqref{wtrace} and~\eqref{wdiv}. Specifically, the left-hand side is
\begin{align}
	\label{DivTS}
	\mathring{\nabla}^a \omega (\tau_{ab}) &= \frac{1}{8 \pi^2} \xi \Big\{K^c_{\phantom{c} c [b} \mathring{\nabla}^a \mathring{\nabla}_{a]} w - K^c_{\phantom{c} (ab)} \mathring{\nabla}^a \mathring{\nabla}_c w \nonumber\\
    &+ \mathring{\nabla}^a K^c_{\phantom{c} c [b} \mathring{\nabla}_{a]} w - \mathring{\nabla}^a K^c_{\phantom{c} (ab)} \mathring{\nabla}_c w \nonumber\\
	&+ [K^d_{\phantom{d} (ab)} K^c_{\phantom{c} cd} - K^d_{\phantom{d} c (a} K^c_{\phantom{c} b) d} - \mathring{\nabla}_b K^c_{\phantom{c} ca}] \mathring{\nabla}^a w \nonumber\\
	&+ [\mathring{\nabla}^a K^d_{\phantom{d} (ab)} K^c_{\phantom{c} cd} + K^d_{\phantom{d} (ab)} \mathring{\nabla}^a K^c_{\phantom{c} cd} \nonumber\\
    &- \mathring{\nabla}^a K^d_{\phantom{d} c (a} K^c_{\phantom{c} b) d} - K^d_{\phantom{d} c (a} \mathring{\nabla}^a K^c_{\phantom{c} b) d} \nonumber\\
	&- \mathring{\nabla}_b K^{c \phantom{[a} a}_{\phantom{c} [a} K^d_{\phantom{d} d] c} - K^{c \phantom{[a} a}_{\phantom{c} [a|} \mathring{\nabla}_b K^d_{\phantom{d} |d] c}] w \Big\} \nonumber\\
    &+ \mathring{\nabla}^a \tilde{\Theta}_{ab} - \frac{1}{4 \pi^2} \mathring{\nabla}_b v_1,
\end{align}
while the right-hand side is
\begin{align}
	\label{DivTS1}
	&(K^c_{\phantom{c} bd} \mathring{\nabla}_a + K^c_{\phantom{c} ab} \mathring{\nabla}_d + \mathring{\nabla}_a K^c_{\phantom{c} bd} + \mathring{\nabla}_d K^c_{\phantom{c} ab}) \omega (\sigma_c^{\phantom{c} ad}) \nonumber\\
    &- (K^d_{\phantom{d} ac} \mathring{\nabla}_d + \mathring{\nabla}_d K^d_{\phantom{d} ac}) \omega (\sigma_b^{\phantom{b} ac}) - \mathring{\nabla}_b K^c_{\phantom{c} ad} \omega (\sigma_c^{\phantom{c} ad}) \nonumber\\
	&= \mathring{\nabla}^a \omega (\tau_{ab}) - \Big(\mathring{\nabla}^a \tilde{\Theta}_{ab} - \frac{1}{4 \pi^2} \mathring{\nabla}_b v_1 \Big) \nonumber\\
	&+ (K^c_{\phantom{c} bd} \mathring{\nabla}_a + K^c_{\phantom{c} ab} \mathring{\nabla}_d + \mathring{\nabla}_a K^c_{\phantom{c} bd} + \mathring{\nabla}_d K^c_{\phantom{c} ab}) \tilde{\Sigma}_c^{\ ad} \nonumber\\
    &- (K^d_{\phantom{d} ac} \mathring{\nabla}_d + \mathring{\nabla}_d K^d_{\phantom{d} ac}) \tilde{\Sigma}_b^{\ ac} - \mathring{\nabla}_b K^c_{\phantom{c} ad} \tilde{\Sigma}_c^{\ ad}.
\end{align}
Combining Eqs.~\eqref{ambTheta} and~\eqref{ambSigma}, we see that the third axiom is satisfied if
\begin{align}
	\label{AREC}
	&\mathring{\nabla}^a \tilde{\Theta}_{ab} - \frac{1}{4 \pi^2} \mathring{\nabla}_b v_1 
    = \nonumber\\
    &(K^c_{\phantom{c} bd} \mathring{\nabla}_a + K^c_{\phantom{c} ab} \mathring{\nabla}_d + \mathring{\nabla}_a K^c_{\phantom{c} bd} + \mathring{\nabla}_d K^c_{\phantom{c} ab}) \tilde{\Sigma}_c^{\phantom{c} ad} \nonumber\\
    &- (K^d_{\phantom{d} ac} \mathring{\nabla}_d + \mathring{\nabla}_d K^d_{\phantom{d} ac}) \tilde{\Sigma}_b^{\phantom{b} ac} - \mathring{\nabla}_b K^c_{\phantom{c} ad} \tilde{\Sigma}_c^{\phantom{c} ad}.
\end{align}
Note that the term \(  g_{ab} v_1 / (4 \pi^2) \) included in Eq.~\eqref{ambTheta} cancels a similar term in Eq.~\eqref{DivTS}.
Hence, the left-hand side of Eq.~\eqref{AREC} corresponds to the divergence of \( \Theta^{M^2}_{ab} + \Theta_{ab} \); consequently, \( \Theta^{M^2}_{ab} + \Theta_{ab} \) and \( {\Sigma^{M^2}}_c^{\phantom{c} ab} + \Sigma_c^{\phantom{c} ab} \) satisfy an equation analogous to that in the third axiom. Moreover, \( \omega (\tau_{ab}) \) can be further simplified by substituting Eq.~\eqref{wtrace} into Eq.~\eqref{TEMECS}, yielding
\begin{align}
	\omega (\tau_{ab}) &= \frac{1}{8 \pi^2} \Bigg\{-w_{ab} + \frac{1}{2} (1 - 2 \xi) \mathring{\nabla}_a \mathring{\nabla}_b w \nonumber\\
    &+ \frac{1}{2} \Bigg(2 \xi - \frac{1}{2} \Bigg) g_{ab} \square w + \xi \mathring{R}_{ab} w \nonumber\\
	&+ \xi \Big[K^c_{\phantom{c} c (a} \mathring{\nabla}_{b)} w - K^c_{\phantom{c} (ab)} \mathring{\nabla}_c w \nonumber\\
    &+ (K^d_{\phantom{d} (ab)} K^c_{\phantom{c} cd} - K^d_{\phantom{d} c (a} K^c_{\phantom{c} b) d}) w + g_{ab} \mathring{\nabla}_c (K^{c \phantom{d} d}_{\phantom{c} d} w) \Big]  \nonumber\\
    &- g_{ab} v_1 \Bigg\}+ \Theta^{M^2}_{ab} + \Theta_{ab}. \label{TEMECS1}
\end{align}
We proceed to construct \( \tilde{\Theta}_{ab} \) and \( \tilde{\Sigma}_c^{\phantom{c} ab} \).

\subsection{Ambiguities} \label{Scale and renormalization ambiguities}

\subsubsection{Scale ambiguities}

To account for the logarithmic singularity in the two-point function, as well as for our arbitrary choice of length scale in the Hadamard biparametrix, we introduce scale ambiguities. We define these objects as
\begin{align}
	\Theta^{M^2}_{ab} &= -\frac{1}{8 \pi^2} \lim_{x' \rightarrow x} \hat{\tau}_{ab} (x, x') V(x, x') \ln (M^2), \label{AETEC}\\
	{\Sigma^{M^2}}_c^{\phantom{c} ab} &= -\frac{1}{8 \pi^2} \lim_{x' \rightarrow x} \hat{\sigma}_c^{\phantom{c} ab} (x, x') V(x, x') \ln (M^2). \label{AESEC}
\end{align}
To find the explicit form of these ambiguities, we could insert Eqs.~\eqref{VSeries} and~\eqref{vnp} into the above expressions and then take the coincidence limit. However, this procedure reduces to the replacements \( w \rightarrow w - v_0 \ln (M^2) \) and \( w_{ab} \rightarrow w_{ab} - ({v_0}_{ab} + g_{ab} v_1) \ln (M^2) \) in Eqs.~\eqref{DSECS} and~\eqref{TEMECS1}, which implies
\begin{align}
		\Theta^{M^2}_{ab} &= -\frac{1}{8 \pi^2} \Bigg\{-({v_0}_{ab} + g_{ab} v_1) + \frac{1}{2} (1 - 2 \xi) \mathring{\nabla}_b \mathring{\nabla}_a v_0 \nonumber\\
    &+ \frac{1}{2} \Bigg(2 \xi - \frac{1}{2} \Bigg) g_{ab} \square v_0 + \xi \mathring{R}_{ab} v_0 \nonumber\\
	&+ \xi [K^c_{\phantom{c} c (a} \mathring{\nabla}_{b)} v_0 - K^c_{\phantom{c} (ab)} \mathring{\nabla}_c v_0 \nonumber\\
    &+ (K^d_{\phantom{d} (ab)} K^c_{\phantom{c} cd} - K^d_{\phantom{d} c (a} K^c_{\phantom{c} b) d}) v_0 \nonumber\\
	&+ g_{ab} \mathring{\nabla}_c (K^{c \phantom{d} d}_{\phantom{c} d} v_0)] \Bigg\} \ln (M^2),\label{AETEC1}\\
	{\Sigma^{M^2}}_c^{\phantom{c} ab} &= -\frac{1}{16 \pi^2} \xi \{\delta^a_{\phantom{a} c} \mathring{\nabla}^b v_0 - g^{ab} \mathring{\nabla}_c v_0 \nonumber\\
    &+ (g^{ab} K^d_{\phantom{d} dc} - \delta^a_{\phantom{a} c} K^{d \phantom{d} b}_{\phantom{d} d} + K_c^{\phantom{c} ba} - K^{b \phantom{c} a}_{\phantom{b} c}) v_0 \} \ln (M^2).\label{AESEC1}
\end{align}

Using Eqs.~\eqref{vtrace} and~\eqref{vdiv}, one can verify that Eq.~\eqref{AREC} is satisfied for \( \Theta^{M^2}_{ab} \) and \( {\Sigma^{M^2}}_c^{\phantom{c} ab} \). In fact, \( \mathring{\nabla}^a \Theta^{M^2}_{ab} \) and \( {\Sigma^{M^2}}_c^{\phantom{c} ab} \) assume the same form as Eqs.~\eqref{DivTS} and~\eqref{DSECS}, respectively, under the replacement \( w \rightarrow w - v_0 \ln (M^2) \). Consequently, Eq.~\eqref{AREC} is automatically satisfied, and the scale ambiguities are consistent with the third axiom.

\subsubsection{Renormalization ambiguities}

The renormalization ambiguities, \( \Theta_{ab} \) and \( \Sigma_c^{\phantom{c} ab} \), can be derived from a renormalization Lagrangian \( \mathcal{L}_{\text{Ren}} = \mathcal{L}_{\text{Ren}} [g^{ab}, K^c_{\phantom{c} ab}] \) by defining  
\begin{align}
	\label{ThetaSigmaRen}
	\Theta_{ab} &= -\frac{2}{\sqrt{-g}} \frac{\delta (\mathcal{L}_{\text{Ren}} \sqrt{-g})}{\delta g^{ab}}, \nonumber\\
    \Sigma_c^{\phantom{c} ab} &= -\frac{1}{\sqrt{-g}} \frac{\delta (\mathcal{L}_{\text{Ren}} \sqrt{-g})}{\delta K^c_{\phantom{c} ab}}.
\end{align}  
We emphasize that this Lagrangian must be purely geometrical, as its role is to produce objects that cancel geometrical quantities, and it must remain regular in the flat-spacetime limit. Moreover, by construction, it does not involve nondynamical fields. Consequently, its associated action is invariant under diffeomorphisms~\cite{Lee,Cristobal}, ensuring that Eq.~\eqref{AREC} is automatically satisfied for \( \Theta_{ab} \) and \( \Sigma_c^{\phantom{c} ab} \). In addition, in four spacetime dimensions, and working in units where \( c = \hbar = 1 \), but without fixing \( G \), \( \mathcal{L}_{\text{Ren}} \) has units of \( \text{length}^{-4} \).

To construct \( \mathcal{L}_{\text{Ren}} \), we employ the formalism of differential forms \cite{Nakahara_2003}. This approach is particularly convenient because the renormalization Lagrangian, \( L_{\text{Ren}} \), is a $4$-form. In this formalism, the dynamical variables are the tetrad $1$-forms, \( e^\mu \), and the spin connection $1$-form, \( \omega^{\mu \nu} \), where Greek indices label Lorentz indices, and abstract spacetime indices are omitted whenever possible, as is customary. The tetrad is related to the spacetime metric via  
\begin{equation}
    g^{ab}\, e_a^{\phantom{a} \mu}\, e_b^{\phantom{b} \nu} = \eta^{\mu\nu},
\end{equation}  
where \( \eta^{\mu\nu} \) is the inverse Minkowski metric and is used to raise Greek indices, while \( \eta_{\mu\nu} \), with identical components, lowers them. Furthermore, \( L_{\text{Ren}} \) is a Lorentz scalar, and to preserve covariance, it can depend on the spin connection only through curvature and torsion~\cite{Cristobal}. The components of these two fields in a coordinate basis have dimensions of \( \text{length}^{-2} \) and \( \text{length}^{-1} \), respectively. Since \( \mathcal{L}_{\text{Ren}} \) is obtained below by taking the Hodge dual of \( L_{\text{Ren}} \), we treat the curvature and torsion 2-forms as having the same dimensions as their corresponding components in a coordinate basis.

Recapitulating, the basic building blocks used to construct \( L_{\text{Ren}} \) are:  
\begin{itemize}
    \item \textbf{Two-forms:} curvature \( R^{\mu \nu} \) and torsion \( T^\mu \), with dimensions \( \text{length}^{-2} \) and \( \text{length}^{-1} \), respectively.  
    \item \textbf{One-forms:} tetrads \( e^\mu \), which are dimensionless.  
    \item \textbf{Zero-forms:} \( \eta_{\mu\nu} \), \( \eta^{\mu\nu} \), and the components of the volume form in the tetrad basis, \( \epsilon_{\mu\nu\rho\sigma} \); all of these objects are dimensionless.  
\end{itemize}  
In addition, the operations we use between forms are:  
\begin{itemize}
    \item \textbf{Wedge product:} for an \( r \)-form \( \alpha \) and an \( m \)-form \( \beta \),
        \begin{align}
            \label{wedge}
            \alpha \wedge \beta &= \sum_{\sigma \in \mathbb{S}^n} \mathrm{sgn}(\sigma)\,
            \alpha_{\mu_{\sigma(1)} \cdots \mu_{\sigma(r)}} 
            \beta_{\mu_{\sigma(r+1)} \cdots \mu_{\sigma(r+m)}} \nonumber\\
            &\quad \times e^{\mu_{\sigma(1)}} \otimes \cdots \otimes e^{\mu_{\sigma(r+m)}},
        \end{align}
        where \( \mathbb{S}^n \) is the permutation group of \( n \) elements. The wedge product combines an \( r \)-form and an \( m \)-form into an \( (r+m) \)-form.  
    \item \textbf{Hodge dual:} defined on a tetrad basis element by
        \begin{align}
            \label{Hodge}
            *(e^{\alpha_1} \wedge \cdots \wedge e^{\alpha_r})
            &= \frac{\epsilon^{\alpha_1 \cdots \alpha_r}_{\phantom{\alpha_1 \cdots \alpha_r} \beta_{r+1} \cdots \beta_m}}{(m - r)!}\,
                       e^{\beta_{r+1}} \wedge \cdots \wedge e^{\beta_m},
        \end{align}
        where \( r \leq 4 \) and \( r + m = 4 \).  
    \item \textbf{Exterior derivative:} denoted by \( \text{d} \), it acts on an \( r \)-form field and produces an \( (r+1) \)-form by taking a partial derivative and antisymmetrizing all indices. It satisfies a graded Leibniz rule.  
\end{itemize}  

Taking into account the identities~\cite{Nakahara_2003}  
\begin{eqnarray}
T^\mu &=& \text{d}e^\mu + \omega^{\mu\nu} \wedge e_\nu ,\\ 
R^\mu_{\ \nu} \wedge e^\nu &=& \text{d}T^\mu + \omega^{\mu\nu}\wedge T_\nu ,\\
0 &=& \text{d}R^{\mu \nu} + \omega^{\mu}_{\phantom{\mu} \rho}\wedge R^{\rho\nu} + \omega^{\nu}_{\phantom{\nu} \rho} \wedge R^{\mu\rho} ,
\end{eqnarray}  
and omitting “topological” terms of the form \( \int \text{d}\alpha \), which do not contribute to \( \Theta_{ab} \) and \( \Sigma_c^{\phantom{c} ab} \), the most general form of \( L_{\text{Ren}} \) with dimensions of \( \text{length}^{-4} \) is  
\begin{align}
    \label{LRenform}
    L_{\text{Ren}} &= \tilde{\alpha}_1 * (R^{\mu \nu} \wedge e^\rho \wedge e^\sigma) R^{\alpha \beta} \wedge e^\gamma \wedge e^\delta \epsilon_{\mu \nu \rho \sigma} \epsilon_{\alpha \beta \gamma \delta} + \tilde{\alpha}_2 * (R^{\mu \nu} \wedge e^\rho \wedge e^\sigma) R_{\mu \nu} \wedge e_\rho \wedge e_\sigma \nonumber\\
    &+ \tilde{\alpha}_3 * (R^{\mu \nu}) \wedge R^{\alpha \beta} \epsilon_{\mu \nu \alpha \beta} + \tilde{\alpha}_4 * (T^\mu \wedge T_\mu) R^{\nu \rho} \wedge e^\alpha \wedge e^\beta \epsilon_{\nu \rho \alpha \beta} + \tilde{\alpha}_5 * (T^\mu \wedge T_\mu) R^{\nu \rho} \wedge e_\nu \wedge e_\rho  \nonumber\\
    &+ \tilde{\alpha}_6 * (T_\mu \wedge T_\nu) R^{\mu \rho} \wedge e^\nu \wedge e_\rho+ \tilde{\alpha}_7 * (T^\mu \wedge T_\mu) T^\nu \wedge T_\nu + \tilde{\alpha}_8 * (T^\mu \wedge T^\nu) T_\mu \wedge T_\nu + \tilde{\beta}_1 R^{\mu \nu} \wedge e_\mu \wedge e_\nu  \nonumber\\
    &+ \tilde{\beta}_2 R^{\mu \nu} \wedge e^\rho \wedge e^\sigma \epsilon_{\mu \nu \rho \sigma}+ \tilde{\gamma} e^\mu \wedge e^\nu \wedge e^\rho \wedge e^\sigma \epsilon_{\mu \nu \rho \sigma},
\end{align}
where \( \tilde{\alpha}_{\text{i}}, \tilde{\beta}_{\text{i}} \), and \( \tilde{\gamma} \) are dimensionless renormalization coupling constants. Observe that the topological terms in this setting are given by~\cite{Mardones_1991}
    \begin{align}
        T^\mu \wedge T_\mu - R^{\mu \nu} e_\mu e_\nu &= \text{d} (e^\mu \wedge T_\mu), \label{NiehYan}\\
        R^{\mu \nu} \wedge R_{\mu \nu} &= \text{d} \Big(\omega_{\mu \nu} \wedge R^{\mu \nu} \nonumber\\
        &+ \frac{1}{3} \omega_{\mu \nu} \wedge \omega^{\nu \rho} \wedge \omega_\rho^{\phantom{\rho} \mu} \Big), \label{Pontryagin}\\
       \epsilon_{\mu \nu \rho \sigma} R^{\mu \nu} \wedge R^{\rho \sigma}  &= \text{d}  \Big[\epsilon_{\mu \nu \rho \sigma} \Big(\omega^{\mu \nu} \wedge R^{\rho \sigma}\nonumber\\
        &+ \frac{1}{3} \omega^{\mu \nu} \wedge \omega^{\rho \kappa} \wedge \omega_\kappa^{\phantom{\kappa} \sigma} \Big) \Big] \label{Euler}.
    \end{align}

To obtain \( \mathcal{L}_{\text{Ren}} \), we apply the Hodge dual to Eq.~\eqref{LRenform}. In terms of the contorsion, it reads  
\begin{align}
    \label{LRen}
        \mathcal{L}_{\text{Ren}} &= \alpha_1 R^2 + \alpha_2 R_{abcd} R^{abcd} + \alpha_3 R^{ab}_{\phantom{ab} ef} R^{cdef} \epsilon_{abcd} \nonumber\\
        &+ \alpha_4 R K_\epsilon + \alpha_5 U^e_{\phantom{e} eabcd} R^{abcd} + \alpha_6 V_{abcd} R^{abcd} \nonumber\\
        &+ \alpha_7 U^e_{\phantom{e} eabcd} K^{fab} K_f^{\phantom{f} cd} + \alpha_8 U^{(ef)}_{\phantom{(ef)} \ abcd} K_e^{\phantom{e} ab} K_f^{\phantom{f} cd} \nonumber\\
        &+ \beta_1 R_{abcd} \epsilon^{abcd} + \beta_2 R + \gamma,
\end{align}
where, for \( [\mathcal{L}_{\text{Ren}}] = \text{length}^{-4} \), the renormalization coupling constants acquire the appropriate dimensions: \( [\alpha_i] = \text{length}^{0} \), \( [\beta_i] = \text{length}^{-2} \), and \( [\gamma] = \text{length}^{-4} \).  
We further define the auxiliary tensors  
\begin{align}
    K_\epsilon &= K^e_{\phantom{e} ab} K_{ecd} \epsilon^{abcd}, \label{Ke}\\
    U^{ef}_{\phantom{ef} abcd} &= K^e_{\phantom{e} [ab]} K^f_{\phantom{f} [cd]} - K^e_{\phantom{e} [ac]} K^f_{\phantom{f} [bd]} + K^e_{\phantom{e} [ad]} K^f_{\phantom{f} [bc]}, \label{Uefabcd}\\
    V^a_{\phantom{a} bcd} &= K^e_{\phantom{e} eb} K^a_{\phantom{a} [cd]} + K^e_{\phantom{e} ec} K^a_{\phantom{a} [bd]} - K^e_{\phantom{e} ed} K^a_{\phantom{a} [bc]} \nonumber\\
    &- 2 \left(K^a_{\phantom{a} [be]} K^e_{\phantom{e} [cd]} + K^a_{\phantom{a} [ce]} K^e_{\phantom{e} [bd]} - K^a_{\phantom{a} [de]} K^e_{\phantom{e} [bc]}\right), \label{Vabcd}
\end{align}
which satisfy \( U^{ef}_{\phantom{ef} abcd} = U^{ef}_{\phantom{ef} [ab] cd} = U^{ef}_{\phantom{ef} ab [cd]} \), \( U^{(ef)}_{\phantom{(ef)} \ abcd} = U^{(ef)}_{\phantom{(ef)} \ cdab} \), and \( V^a_{\phantom{a} bcd} = V^a_{\phantom{a} b [cd]} \).  

We present the contributions of each term to \( \Theta_{ab} \) and \( \Sigma_c^{\phantom{c} ab} \) in the following tables, where each row corresponds to a single renormalization term, indicated by its coupling constant.
\begin{center}
    \begin{table}[H]
        \begin{tabular}{p{0.03 \textwidth}|p{0.534 \textwidth}|p{0.43 \textwidth}}
        & \begin{center} \( \Theta_{ab} \) \end{center} & \begin{center} \( \Sigma_c^{\phantom{c} ab} \) \end{center}\\
            \hline
                \( \alpha_1 \) & \makecell{\\ \( \begin{aligned} &-2 \Big\{ 2 R \mathring{R}_{ab} - 2 \mathring{\nabla}_a \mathring{\nabla}_b R + 2 \square R g_{ab} - \frac{1}{2} R^2 g_{ab}\\ &\qquad + 2(K^c_{\phantom{c} c (a} \mathring{\nabla}_{b)} R - K^c_{\phantom{c} (ab)} \mathring{\nabla}_c R)\\ &\qquad + 2R (K^d_{\phantom{d} (ab)} K^c_{\phantom{c} cd} - K^d_{\phantom{d} c (a} K^c_{\phantom{c} b) d}) + 2 \mathring{\nabla}_c (R K^{c \phantom{d} d}_{\phantom{c} d}) g_{ab} \Big\} \end{aligned} \)\\\\} & \makecell{\( \begin{aligned} &-2 \{\delta^a_{\phantom{a} c} \mathring{\nabla}^b R - g^{ab} \mathring{\nabla}_c R\\ &\qquad + R (g^{ab} K^d_{\phantom{d} dc} - \delta^a_{\phantom{a} c} K^{d \phantom{d} b}_{\phantom{d} d} + K_c^{\phantom{c} ba} - K^{b \phantom{c} a}_{\phantom{b} c}) \} \end{aligned} \)}\\
                \hline
                \( \alpha_2 \) & \makecell{\\ \( \begin{aligned} &-2 \Big\{2 R_{acde} R_b^{\phantom{b} cde} - \frac{1}{2} R_{cdef} R^{cdef} g_{ab} + 4 \mathring{\nabla}_d \mathring{\nabla}_c R_{(a \phantom{c} b)}^{\phantom{(a} c \phantom{b)} d}\\ &\qquad + 4 \mathring{\nabla}_e (R_{(a \phantom{c} b)}^{\phantom{(a} c \phantom{b)} d} K^e_{\phantom{e} cd} + R_{(a}^{\phantom{(a} cde} K_{b) cd}) \Big\} \end{aligned} \)\\\\} & \makecell{\( \begin{aligned} &-4 \{ -\mathring{\nabla}_d R^{adb}_{\phantom{adb} c} + R^{adb}_{\phantom{adb} e} K^e_{\phantom{e} dc} + R^{ad \phantom{c} e}_{\phantom{ad} c} K^b_{\phantom{b} de} \} \end{aligned} \)}\\
                \hline
                \( \alpha_3 \) & \makecell{\\ \( \begin{aligned} &-2 \{4R_{(a|}^{\phantom{(a|} cf \ell} R^{de}_{\phantom{de} f \ell} \epsilon_{|b) cde} - R_{cd \ell m} R_{ef}^{\phantom{ef} \ell m} \epsilon^{cdef} g_{ab}\\ &\qquad + 4 \mathring{\nabla}_d \mathring{\nabla}_c R^{ef \phantom{(a} d}_{\phantom{ef} (a} \epsilon_{b) \phantom{c} ef}^{\phantom{b)} c}\\ &\qquad + 4 \mathring{\nabla}_e (R^{f \ell \phantom{(a|} d}_{\phantom{ef} (a|} K^e_{\phantom{e} cd} + R^{f \ell de} K_{(a| cd}) \epsilon_{|b) \phantom{c} f \ell}^{\phantom{|b)} c} \} \end{aligned} \)\\\\} & \makecell{\( \begin{aligned} &-4 \{-\mathring{\nabla}_d R^{efb}_{\phantom{efb} c} \epsilon^{ad}_{\phantom{ad} ef} + (R^{f \ell b}_{\phantom{f \ell b} e} K^e_{\phantom{e} dc} + R^{f \ell \phantom{c} e}_{\phantom{f \ell} c} K^b_{\phantom{b} de}) \epsilon^{ad}_{\phantom{ad} f \ell} \} \end{aligned} \)}\\
                \hline
                \( \alpha_4 \) & \makecell{\\ \( \begin{aligned} &-2 \{R[2(K^f_{\phantom{f} (a| c} - K^f_{\phantom{f} c (a|})K_{fde} \epsilon_{|b)}^{\quad cde} - K_{(a| cd} K_{|b) ef} \epsilon^{cdef}]\\ &\qquad + K_\epsilon \mathring{R}_{ab} - \mathring{\nabla}_a \mathring{\nabla}_b K_\epsilon + \square K_\epsilon g_{ab} - RK_\epsilon g_{ab}\\ &\qquad + K^c_{\phantom{c} c (a} \mathring{\nabla}_{b)} K_\epsilon - K^c_{\phantom{c} (ab)} \mathring{\nabla}_c K_\epsilon\\ &\qquad + K_\epsilon (K^d_{\phantom{d} (ab)} K^c_{\phantom{c} cd} - K^d_{\phantom{d} c (a} K^c_{\phantom{c} b) d}) + \mathring{\nabla}_c (K_\epsilon K^{c \phantom{d} d}_{\phantom{c} d}) g_{ab} \}  \end{aligned} \)\\\\} & \makecell{\( \begin{aligned} &-\{\delta^a_{\phantom{a} c} \mathring{\nabla}^b K_\epsilon - g^{ab} \mathring{\nabla}_c K_\epsilon\\ &\quad \ + K_\epsilon (g^{ab} K^d_{\phantom{d} dc} - \delta^a_{\phantom{a} c} K^{d \phantom{d} b}_{\phantom{d} d} + K_c^{\phantom{c} ba} - K^{b \phantom{c} a}_{\phantom{b} c})\\ &\quad \ + 2R g_{cf} K^{[f| de} \epsilon^{a |b] de} \} \end{aligned} \)}\\
                      \end{tabular}
   \end{table}
  \begin{table}[H]
     \begin{tabular}{p{0.03 \textwidth}|p{0.5 \textwidth}|p{0.45 \textwidth}}
    & \begin{center} \( \Theta_{ab} \) \end{center} & \begin{center} \( \sigma_c^{\phantom{c} ab} \) \end{center}\\
     \hline
                \( \alpha_5 \) & \makecell{\\ \( \begin{aligned} &-2 \Big\{2U^f_{\phantom{f} f (a| cde} R_{|b)}^{\phantom{|b)} cde} + U^f_{\phantom{f} fcde (a|} R^{cde}_{\phantom{cde} |b)}\\ &\qquad - U_{(ab) cdef} R^{cdef} g_{ab} - \frac{1}{2} U^\ell_{\phantom{\ell} \ell cdef} R^{cdef} g_{ab}\\ &\qquad + 2[\mathring{\nabla}_d \mathring{\nabla}_c U^{e \phantom{e (a} c \phantom{b)} d}_{\phantom{e} e (a \phantom{c} b)}\\ &\qquad \quad \ \ + \mathring{\nabla}_e (U^{f \phantom{f (a} c \phantom{b)} d}_{\phantom{f} f (a \phantom{c} b)} K^e_{\phantom{e} cd} + U^{f \phantom{f (a} cde}_{\phantom{f} f (a} K_{b) cd})] \Big\} \end{aligned} \)\\\\} & \makecell{\( \begin{aligned} &-2 \{-\mathring{\nabla}_d U^{e \phantom{e} adb}_{\phantom{e} e \phantom{adb} c} + U^{f \phantom{f} adb}_{\phantom{f} f \phantom{adb} e} K^e_{\phantom{e} dc} + U^{f \phantom{f} ad \phantom{c} e}_{\phantom{f} f \phantom{ad} c} K^b_{\phantom{b} de}\\ &\qquad + 2g_{cf} (R^{a [b| de} + R^{dea [b|} + R^{ae [b| d} \\ &\qquad \qquad \quad - R^{[b| ead} - R^{ad [b| e} + R^{[b| dae}) K^{|f]}_{\phantom{|f]} de} \} \end{aligned} \)}\\
                \hline
                \( \alpha_6 \) & \makecell{\( \begin{aligned} &-2 \Big\{-V^{c \phantom{(a|} d}_{\phantom{c} (a| \phantom{d} e} R_{|b) cd}^{\phantom{|b) cd} e} + V^{cd}_{\phantom{cd} (a| e} R_{cd |b)}^{\phantom{cd |b)} e} \\ &\qquad - \frac{1}{2} V_{cdef} R^{cdef} g_{ab}+ \mathring{\nabla}_d \mathring{\nabla}_c (V_{(a \phantom{c} b)}^{\phantom{(a} c \phantom{b)} d} + V^{c \phantom{(a} d}_{\phantom{c} (a \phantom{d} b)})\\ &\qquad + \mathring{\nabla}_e [(V_{(a \phantom{c} b)}^{\phantom{(a} c \phantom{b)} d} + V^{c \phantom{(a} d}_{\phantom{c} (a \phantom{d} b)}) K^e_{\phantom{e} cd}\\ &\qquad + (V_{(a}^{\phantom{(a} cde} - V^{c \phantom{(a} de}_{\phantom{c} (a}) K_{b) cd}] \Big\} \end{aligned} \)} & \makecell{\\ \( \begin{aligned} &-\Big\{2[-\mathring{\nabla}_d V^{[ad]b}_{\phantom{[ad] b} c} + V^{[ad]b}_{\phantom{[ad] b} e} K^e_{\phantom{e} dc} + V^{[ad] \phantom{c} e}_{\phantom{[ad]} c} K^b_{\phantom{b} de}]\\ &\quad + \Big[g^{ab} \Big(\frac{1}{2} R_{cd}^{\phantom{cd} ef} - R_{d \phantom{[e} c}^{\phantom{d} [e \phantom{c} f]} \Big) \\ &\qquad - \delta^a_{\phantom{a} c} \Big(\frac{1}{2} R^{b \phantom{d} ef}_{\phantom{b} d} - R_d^{\phantom{d} [e|b|f]} \Big) \Big] K^d_{\phantom{d} ef}\\ &\quad + \Big[-\frac{1}{2} \Big(R^{[a \phantom{d} b]}_{\phantom{[a} d \phantom{b]} c} + R_{cd}^{\phantom{cd} ab} \Big) + R^{ab}_{\phantom{ab} cd} - R^{[a \phantom{c} b]}_{\phantom{[a} c \phantom{b]} d} \Big] K^{d \phantom{e} e}_{\phantom{d} e}\\ &\quad + (R^{[a| \phantom{c} de}_{\phantom{[a|} c} - 2R_c^{\phantom{c} d [a| e}) K^{|b]}_{\phantom{|b]} [de]} \\ &\qquad - (R^{abde} + 2R^{[a|d|b] e}) K_{c [de]}\\ &\quad - (R^{de [a}_{\phantom{de [a} [c} + 2R^{[a| d \phantom{[c} e}_{\phantom{[a| d} [c}) K^{|b]}_{\phantom{|b]} e] d} \\ &\qquad - (R^{deab} + 2R^{[a|d|b] e}) K_{d [ec]} \Big\} \end{aligned} \)\\\\}\\
                \hline
                \( \alpha_7 \) & \makecell{\\ \(\begin{aligned} &-2 \Big\{-U_{(ab) cdef} K^{\ell cd} K_\ell^{\phantom{\ell} ef} - U^\ell_{\phantom{\ell} \ell cdef} K_{(a}^{\phantom{(a} cd} K_{b)}^{\phantom{b)} ef}\\ &\qquad + U^f_{\phantom{f} f (a| cde} (K^{\ell \phantom{|b)} c}_{\phantom{\ell} |b)} - K^{\ell c}_{\phantom{\ell c} |b)}) K_\ell^{\phantom{\ell} de}\\
                &\qquad + U^f_{\phantom{f} fcd (a| e} K^{\ell cd} (K_{\ell |b)}^{\phantom{\ell |b)} e} - K_{\ell \phantom{e} |b)}^{\phantom{\ell} e})\\
                &\qquad - \frac{1}{2} U^\ell_{\phantom{\ell} \ell cdef} K^{mcd} K_m^{\phantom{m} ef} g_{ab} \Big\} \end{aligned} \)\\\\} & \makecell{\( -6g_{c \ell} U^{f \phantom{f} a [b| de}_{\phantom{f} f} K^{|\ell]}_{\phantom{|\ell]} de} \)}\\
                \hline
                \( \alpha_8 \) & \makecell{\\ \( \begin{aligned} &-2 \Big\{-g_{(a| m} (U_{|b) \ell cdef} + U_{\ell |b) cdef}) K^{(\ell| cd} K^{|m) ef}\\ &\qquad + U^{(f \ell)}_{\phantom{(f \ell)} \ (a| cde} (K_{f |b)}^{\phantom{f |b)} c} - K_{f \phantom{c} |b)}^{\phantom{f} c}) K_\ell^{\phantom{\ell} de}\\
                &\qquad + U^{(f \ell)}_{\phantom{(f \ell)} \ cd (a| e} K_f^{\phantom{f} cd} (K_{\ell |b)}^{\phantom{\ell |b)} e} - K_{\ell \phantom{e} |b)}^{\phantom{\ell} e})\\
                &\qquad - \frac{1}{2} U^{(\ell m)}_{\phantom{(\ell m)} cdef} K_\ell^{\phantom{\ell} cd} K_m^{\phantom{m} ef} g_{ab} \Big\} \end{aligned} \)\\\\} & \makecell{\( -6(U_{(cf)}^{\phantom{(cf)} abde} - U^{b \phantom{(c} a \phantom{f)} de}_{\phantom{b} (c \phantom{a} f)}) K^f_{\phantom{f} de} \)}
        \end{tabular}
    \end{table}
    
 \begin{table}[H]
        \begin{tabular}{p{0.03 \textwidth}|p{0.52\textwidth}|p{0.43 \textwidth}}
        & \begin{center} \( \Theta_{ab} \) \end{center} & \begin{center} \( \sigma_c^{\phantom{c} ab} \) \end{center}\\
                \hline
                \( \beta_1 \) & \makecell{\\ \( -4 \{(2K^f_{\phantom{f} (a| d} - K^f_{\phantom{F} d (a|}) K_{ecf} \epsilon_{|b)}^{\phantom{|b)} cde} - K^\ell_{\phantom{\ell} ce} K_{fd \ell} \epsilon^{cdef} g_{ab} \} \)\\\\} & \makecell{\( -2 \{\epsilon^{adb}_{\phantom{adb} e} K^e_{\phantom{e} dc} + \epsilon^{ad \phantom{c} e}_{\phantom{ad} c} K^b_{\phantom{b} de} \} \)}\\
                \hline
                \( \beta_2 \) & \makecell{\\ \( -2 \{\mathring{G}_{ab} + K^d_{\phantom{d} (ab)} K^c_{\phantom{c} cd} - K^d_{\phantom{d} c (a} K^c_{\phantom{c} b) d} - K^{d \phantom{[c} c}_{\phantom{d} [c} K^e_{\phantom{e} e] d} g_{ab} \} \)\\\\} & \makecell{\( -\{g^{ab} K^d_{\phantom{d} dc} - \delta^a_{\phantom{a} c} K^{d \phantom{d} b}_{\phantom{d} d} + K_c^{\phantom{c} ba} - K^{b \phantom{c} a}_{\phantom{b} c} \} \)}\\
                \hline
                \( \gamma \) & \makecell{\\ \( g_{ab} \)\\\\} & \makecell{\( 0 \)}
        \end{tabular}
    \end{table}
\end{center}

From inspecting Eq.~\eqref{LRen}, it follows that \( \beta_2 \) renormalizes Newton’s constant, while \( \gamma \) plays the role of a cosmological constant (which is known to have relevant applications in GR~\cite{JuarezDeSitter,JuarezDeSitter2}). Furthermore, in semiclassical GR with a nonminimally coupled Klein--Gordon field, only the terms associated with \( \alpha_1, \ \alpha_2, \ \beta_2 \), and \( \gamma \) contribute \cite{Decanini_2008}. This is consistent with our findings, since the only terms with a nontrivial dependence on \( K^c_{\phantom{c} ab} \) are those coupled to these constants, along with \( \beta_1 \) and \( \alpha_3 \). However, the \( \beta_1 \) term is the Holst term, which vanishes in GR, whereas the term coupled to \( \alpha_3 \) is a boundary term in GR. In particular, the Hodge dual of Eq.~\eqref{Pontryagin} is proportional to \( R^{\mu \nu}_{\phantom{\mu \nu} \alpha \beta} R_{\mu \nu \gamma \delta} \epsilon^{\alpha \beta \gamma \delta} \), while the dual of the term coupled to \( \tilde{\alpha}_3 \) is \( R_{\alpha \beta}^{\phantom{\alpha \beta} \mu \nu} R_{\gamma \delta \mu \nu} \epsilon^{\alpha \beta \gamma \delta} \). In GR, these expressions coincide; however, this equivalence no longer holds in the presence of torsion.

Lastly, substituting Eqs.~\eqref{DSECS} and~\eqref{TEMECS1} into the equations of motion reveals that the semiclassical equations differ significantly from their classical counterparts. In the metric equation, both scaling and renormalization ambiguities give rise to fourth-order derivative terms, which complicate the analysis of the Cauchy problem~\cite{Cauchy2020,Cauchy2022,Cauchy2023,Cauchy2023-1,Cauchy2024}. In addition, semiclassically, the contorsion equation ceases to be purely algebraic. Consequently, in a semiclassical treatment, it may occur that \( K^c_{\phantom{c} ab} \neq 0 \) in regions where \( \omega(\sigma_c^{\phantom{c} ab}) = 0 \). A detailed analysis of this result is left for future work.

\subsection{Conformal anomaly} \label{Conformal anomaly}

Lastly, we compute \( \omega (\tau^a_{\phantom{a} a}) \), ignoring renormalization ambiguities, for simplicity, to compare it with the classical expression \( \tau^a_{\phantom{a} a} \). From Eqs.~\eqref{wtrace},~\eqref{vtrace},~\eqref{TEMECS1}, and~\eqref{AETEC1}, and for the parameters that make the classical theory conformally invariant, \( m = 0 \) and \( \xi = 1/6 \), we find
\begin{equation}
    \label{Trtauren1}
    \omega (\tau^a_{\phantom{a} a}) =
    \frac{1}{4 \pi^2}\, \mathring{\nabla}_c \!\left\{\frac{1}{3} g^{ab} K^c_{\phantom{c} ab} [w - v_0 \ln (M^2)] \right\}
    + \frac{v_1}{4 \pi^2}.
\end{equation}
The first term is still a total divergence; however, the presence of the term \( v_1 / (4 \pi^2) \) spoils this symmetry. This implies the existence of a conformal anomaly.

Additionally, only the last three terms of \( v_1 \) are nonvanishing or a total divergence [see Eq.~\eqref{v1}]. The first of these terms is purely torsional, while the remaining two are purely metric and coincide with those reported in Ref.~\cite{Decanini_2008}. From the inspection of these contributions, we conclude that the conformal anomaly is, in general, not removed by torsion. In this sense, the conformal anomaly appears to be an intrinsic feature of the semiclassical formalism.

\section{Conclusions}\label{Conc}

In this work, we show that the Hadamard renormalization procedure for a nonminimally coupled, free, massive Klein--Gordon field can be extended to spacetimes with torsion, at least within Einstein--Cartan theory. The axiomatic framework we propose is a direct generalization of Wald's axioms and proves sufficient to establish a well-defined renormalization scheme. Using Hadamard states, we regularize \( \omega(\tau_{ab}) \) and \( \omega(\sigma_c^{\phantom{c} ab}) \) by subtracting their divergence structure. This subtraction naturally introduces scale and renormalization ambiguities which do not depend on the state and can be explicitly computed.

Although the theory presented here differs fundamentally from semiclassical GR, the third axiom can still be motivated by the corresponding classical expression. Notably, the term \( g_{ab} v_1 / (4\pi^2) \) must be added to \( \omega(\tau_{ab}) \) regardless of whether torsion is present, even though its specific form depends on the theory under consideration.

Differential forms prove particularly useful in constructing the renormalization Lagrangian, which we can obtain from a Lorentz scalar, covariant $4$-form with units of \( \text{length}^{-4} \), that is regular when spacetime is flat. This formalism also allows us to identify topological terms, which do not contribute to renormalization. We further show that the conformal anomaly, ignoring renormalization ambiguities, persists in the presence of torsion and remains driven by the \( v_1 \) term, as in semiclassical GR.

A key point enabling Hadamard renormalization is that the matter sector is free, i.e., the classical action is quadratic in the matter field. Consequently, the expectation values of the energy--momentum and spin--density operators can be computed directly from the renormalized two-point function. It remains to be seen whether a Hadamard-like renormalization procedure is adequate for actions containing higher-order terms, which would require going beyond the two-point function. These aspects may become more accessible with torsion, as one could construct theories in which the energy--momentum tensor remains quadratic in the matter fields while the spin--density tensor contains higher-order terms.

Moreover, the semiclassical theory provides a qualitatively different physical description compared to the classical one. In Einstein--Cartan theory, the equation of motion for \( K^c_{\phantom{c} ab} \) is algebraic, so this tensor is only nonvanishing in regions where \( \sigma_c^{\phantom{c} ab} \neq 0 \). However, the renormalization ambiguities modify the contorsion equation, turning it into a differential rather than an algebraic equation. It remains an open question whether this feature allows for the detection of torsion outside polarized matter. Still, one of the main takeaways of this work is that semiclassical effects provide access to a rich phenomenology within modified gravity theories that could otherwise remain inaccessible. Therefore, the semiclassical framework serves as a valuable avenue for exploring such theories.

Finally, a subtlety concerns the choice of geodesic distance in the Hadamard bi-parametrix. Since geodesics and autoparallel curves generally do not coincide, one can alternatively employ the autoparallel distance in the regularization scheme. Whether this choice leads to distinct physical predictions or merely equivalent formulations remains an open question.

\section*{Acknowledgments}
We acknowledge valuable feedback from Daniel Sudarsky. This work was supported by the UNAM-DGAPA-PAPIIT grant IN101724 and by the SECIHTI through its graduate school scholarship program.

\appendix

\section{Geometrical bitensors} \label{Appendix}

In this appendix, several relevant expressions concerning geometrical bitensors are presented, which we use throughout the text. Their proofs can be found in Refs.
~\cite{Christensen_1976,Decanini_2006,Decanini_2008,Poisson}. We begin by listing properties of $\sigma(x,x')$, a biscalar, i.e., a scalar that depends on two points, which corresponds to half the squared geodesic distance between $x$ and $x'$. For $\sigma(x,x')$ to be well-defined, it is assumed that its arguments lie within a normal convex hull. One can show that
\begin{equation}
    2 \sigma = \mathring{\nabla}^a \sigma\, \mathring{\nabla}_a \sigma. \label{PropSig1}
\end{equation}
In addition, when $\sigma(x,x')$ is small, one can prove that
\begin{align}
    \mathring{\nabla}_b \mathring{\nabla}_a \sigma &= g_{ab}
        - \frac{1}{3}\, \mathring{R}_{ac_1 bc_2}\, \mathring{\nabla}^{c_1} \sigma\, \mathring{\nabla}^{c_2} \sigma
        + \mathcal{O} (\sigma^{3/2}), \label{PropSig3}\\
    g_b^{\phantom{b} b'}\, \mathring{\nabla}_{b'} \mathring{\nabla}_a \sigma &= -g_{ab}
        - \frac{1}{6}\, \mathring{R}_{ac_1 bc_2}\, \mathring{\nabla}^{c_1} \sigma\, \mathring{\nabla}^{c_2} \sigma
        + \mathcal{O} (\sigma^{3/2}), \label{PropSig4}
\end{align}
where unprimed indices are associated with $x$ and primed indices with $x'$. In addition, \( g_a^{\phantom{a} a'} \) is the bitensor that parallel-transports vectors from \( x \) to \( x' \) along the geodesic connecting them. This bitensor satisfies
\begin{eqnarray}
    \mathring{\nabla}_c g_{ab'}\, \mathring{\nabla}^c \sigma &=& 0, \label{Propg1}\\
    \lim_{x' \rightarrow x} g_{ab'} &=& g_{ab}. \label{Propg2}
\end{eqnarray}
Further details on \( g_a^{\phantom{a} a'} \) can be found in Refs.~\cite{Brown_1986,Decanini_2008}.

When taking the coincidence limit, one can show that \( \sigma \) satisfies the following properties:
\begin{align} 
    \lim_{x' \rightarrow x} \mathring{\nabla}_a \sigma &= 0=\lim_{x' \rightarrow x} \mathring{\nabla}_{a'} \sigma, \label{Propsig5}\\
    \lim_{x' \rightarrow x} \mathring{\nabla}_c \mathring{\nabla}_b \mathring{\nabla}_a \sigma &= 0=\lim_{x' \rightarrow x} \mathring{\nabla}_c \mathring{\nabla}_b \mathring{\nabla}_{a'} \sigma , \label{PropSig6}\\
    \lim_{x' \rightarrow x} \mathring{\nabla}_d \mathring{\nabla}_c \mathring{\nabla}_b \mathring{\nabla}_a \sigma
        &= -\frac{1}{3}\, (\mathring{R}_{acbd} + \mathring{R}_{adbc}), \label{PropSig7}\\
    \lim_{x' \rightarrow x} \mathring{\nabla}_d \mathring{\nabla}_c \mathring{\nabla}_b \mathring{\nabla}_{a'} \sigma
        &= \frac{1}{2}\, (\mathring{R}_{abcd} - \mathring{R}_{acdb} + \mathring{R}_{adbc}) \nonumber\\
        &+ \frac{1}{3}\, (\mathring{R}_{acbd} + \mathring{R}_{adbc}). \label{PropSig10}
\end{align}

Furthermore, the Van~Vleck--Morette determinant is defined by
\begin{equation}
    \label{VMD}
    \Delta (x, x') = 
    \frac{\det[-\mathring{\nabla}_{\mu} \mathring{\nabla}_{\nu'} \sigma (x, x')]}
    {\sqrt{-g(x)}\, \sqrt{-g(x')}},
\end{equation}
and satisfies \( \Delta (x, x') \rightarrow 1 \) in the coincidence limit. This object appears in the following expression:
\begin{equation}
        \square_x \sigma = 4 - 2\, \Delta^{-1/2}\, \mathring{\nabla}_a (\Delta^{1/2} \mathring{\nabla}^a \sigma), \label{PropSig2}
        \end{equation}
    which is valid for any pair of points \( x \) and \( x' \). In the particular cases where $\sigma(x,x')$ is small, the Van~Vleck--Morette determinant, and quantities related to it, can be expanded as
    \begin{align}
    0 &= -\Delta^{1/2} + 1 + \frac{1}{12} \mathring{R}_{ab} \mathring{\nabla}^a \sigma \mathring{\nabla}^b \sigma + \mathcal{O} (\sigma^{3/2}), \label{VMDE}\\
    0 &= -\square \Delta^{1/2} + \frac{1}{6} \mathring{R} + \Big(\frac{1}{40} \square \mathring{R}_{ab} - \frac{1}{120} \mathring{\nabla}_a \mathring{\nabla}_b \mathring{R}\nonumber\\
        & + \frac{1}{72} \mathring{R} \mathring{R}_{ab}- \frac{1}{30} \mathring{R}_a^{\phantom{a} c} \mathring{R}_{bc} + \frac{1}{60} \mathring{R}^{cd} \mathring{R}_{acbd} 
        \nonumber\\
        & + \frac{1}{60} \mathring{R}_a^{\phantom{a} cde} \mathring{R}_{bcde} \Big) \mathring{\nabla}^a \sigma \mathring{\nabla}^b \sigma  + \mathcal{O} (\sigma^{3/2}), \label{DVMDE}\\
    0 &= -\Delta^{-1/2} \mathring{\nabla}_a \Delta^{1/2} \mathring{\nabla}^a \sigma + \frac{1}{6} \mathring{R}_{ab} \mathring{\nabla}^a \sigma \mathring{\nabla}^b \sigma \nonumber\\
    &+ \mathcal{O} (\sigma^{3/2}). \label{DDSigE}
\end{align}
To summarize, in this appendix we list some identities used throughout the text. These results provide the necessary tools for the computations presented in the main sections above.

\end{document}